\documentclass[fleqn,usenatbib]{mnras}

\usepackage[T1]{fontenc}

\DeclareRobustCommand{\VAN}[3]{#2}
\let\VANthebibliography\thebibliography
\def\thebibliography{\DeclareRobustCommand{\VAN}[3]{##3}\VANthebibliography}

\usepackage{graphicx}	% Including figure files
\usepackage{amsmath}	% Advanced maths commands

\usepackage{amssymb}	% Extra maths symbols

\usepackage{color,soul} % Needed for \hl{...} command to highlight text
\usepackage{booktabs}
\usepackage[flushleft]{threeparttable}
\usepackage{multirow}

\newcommand{\oi}{O\,{\textsc i}}
\newcommand{\oii}{O\,{\textsc{ii}}}
\newcommand{\nii}{Ni\,{\textsc i}}
\newcommand{\niii}{Ni\,{\textsc{ii}}}

\newcommand{\Ot}[5]{\mbox{$#1\,^#2{\rm #3}^{{\rm #4}}_{\rm #5}$}}
\newcommand{\Nt}[5]{\mbox{$#1\,^#2{\rm #3}^{{\rm #4}}_{\rm #5}$}}

\newcommand{\teff}{T_{\rm eff}}

\newcommand{\Elow}{E_{\rm low}}
\newcommand{\Eup}{E_{\rm up}}

\newcommand{\opd}{\log \tau_{\rm 500}}

\newcommand{\tda}{$\langle{\rm 3D}\rangle$}

\newcommand{\Angstrem}{Å}

\newcommand{\logo}{\log{\rm A(O)}}
\newcommand{\logni}{\log{\rm A(Ni)}}
\title[Solar oxygen abundance]{Solar oxygen abundance}
\author[Bergemann et al.]{
Maria Bergemann,$^{1,2,3}$\thanks{E-mail: bergemann@mpia.de}
Richard Hoppe,$^{1,2,4}$
Ekaterina Semenova, $^{2}$
Mats Carlsson, $^{5,6}$
Svetlana A. Yakovleva,$^{7}$
\newauthor
Yaroslav V. Voronov,$^{7}$
Manuel Bautista,$^{8}$
Ahmad Nemer,$^{9}$
Andrey K. Belyaev,$^{7}$
Jorrit Leenaarts,$^{10}$
\newauthor
Lyudmila Mashonkina,$^{11}$
Ansgar Reiners,$^{12}$ 
Monika Ellwarth$^{12}$ 
\\
\\
$^{1}$ The first two authors have contributed equally to the work presented in this paper \\
$^{2}$ Max Planck Institute for Astronomy, 69117, Heidelberg, Germany \\
$^{3}$ Niels Bohr International Academy, Niels Bohr Institute, Blegdamsvej 17, DK-2100 Copenhagen Ø, Denmark \\
$^{4}$ Ruprecht Karls University, Grabengasse 1, 69117 Heidelberg, Germany \\
$^{5}$ Rosseland Centre for Solar Physics, University of Oslo, P.O. Box 1029 Blindern, NO-0315 Oslo, Norway \\
$^{6}$ Institute of Theoretical Astrophysics, University of Oslo, P.O. Box 1029 Blindern, NO-0315 Oslo, Norway \\
$^{7}$ Department of Theoretical Physics and Astronomy, Herzen University, St. Petersburg, 191186, Russia \\
$^{8}$ Department of Physics, Western Michigan University, Kalamazoo, Michigan, 49008, USA \\
$^{9}$ Princeton University, Department of astrophysical sciences, Princeton, New Jersey, USA \\
$^{10}$ Institute for Solar Physics, Department of Astronomy, Stockholm University, AlbaNova University Centre, SE-106 91 Stockholm, Sweden \\
$^{11}$ Institute of Astronomy of the Russian Academy of Sciences, Pyatnitskaya st. 48, 119017, Moscow, Russia \\
$^{12}$ Institut für Astrophysik, Georg-August-Universität G\"{o}ttingen, Friedrich-Hund-Platz 1, D-37077 G\"{o}ttingen, Germany}

\date{Accepted XXX. Received YYY; in original form ZZZ}

\pubyear{2015}

\begin{document}
\label{firstpage}
\pagerange{\pageref{firstpage}--\pageref{lastpage}}
\maketitle

\begin{abstract}
Motivated by the controversy over the surface metallicity of the Sun, we present a re-analysis of the solar photospheric oxygen (O) abundance. New atomic models of O and Ni are used to perform Non-Local Thermodynamic Equilibrium (NLTE) calculations with 1D hydrostatic (MARCS) and 3D hydrodynamical (Stagger and Bifrost) models. The Bifrost 3D MHD simulations are used to quantify the influence of the chromosphere. We compare the 3D NLTE line profiles with new high-resolution, R $\approx 700\,000$, spatially-resolved spectra of the Sun obtained using the IAG FTS instrument. We find that the O I lines at 777~nm yield the abundance of $\logo = 8.74 \pm 0.03$ dex, which depends on the choice of the H-impact collisional data and oscillator strengths. The forbidden [O~I] line at 630~nm is less model-dependent, as it forms nearly in LTE and is only weakly sensitive to convection. However, the oscillator strength for this transition is more uncertain than for the 777 nm lines. Modelled in 3D NLTE with the Ni I blend, the 630~nm line yields an abundance of $\logo = 8.77 \pm 0.05$ dex. We compare our results with previous estimates in the literature and draw a conclusion on the most likely value of the solar photospheric O abundance, which we estimate at $\logo = 8.75 \pm 0.03$ dex.
\end{abstract}


%%%%%%%%%%%%%%%%%%%%%%%%%%%%%%%%%%%%%%%%%%%%%%%%%%

%%%%%%%%%%%%%%%%% BODY OF PAPER %%%%%%%%%%%%%%%%%%

\section{Introduction}
Oxygen is (behind H and He) the most abundant chemical element in the Universe and it is of major relevance in modern astrophysics, across difference fields,  including precision stellar physics, extragalactic astronomy, planet formation,  and galaxy evolution. Most importantly, oxygen determines much of the opacity in the solar interior \citep[e.g.][]{Bahcall2005,Serenelli2009,Pinsonneault2009}, therefore its abundance is critical to the calculation of Standard Solar Models (SSM), which describe the evolution of the Sun from the pre-main-sequence to the present age of 4.5 Gyr. Oxygen is also the key element in gas-phase spectroscopic diagnostics on extragalactic scales, in particular to infer the metallicities from the H II regions \citep[e.g.][]{Kewley2008,Moustakas2010} - a broadly-used technique that has been applied to many star-forming galaxies to establish their metal-content and to quantify the mass-metallicity relationship across the entire mass range of galaxies in the local universe. Oxygen is also one of the two - in addition to Mg - most common tracers of nucleosynthesis in massive stars, which explode as core-collapse supernovae. Therefore, O abundance is  traditionally used in combination with Fe to map the chemical enrichment and star formation history of stellar populations in the Milky Way and its satellite galaxies \citep[e.g.][]{Tolstoy2009,Barbuy2018}. 

This study focuses on the first problem - the chemical abundance of oxygen in the Sun. Over the past decades, several groups approached this problem from various angles. Some groups attempted to measure the O abundances from the solar photospheric spectrum \citep[e.g.][]{Grevesse1998, Allende2001, Asplund2004, Caffau2008, Pereira2009b, Sitnova2013, Caffau2015, Socas2015, Steffen2015, Amarsi2018}. Other groups derived the solar O abundances from the high-ionised lines in the solar wind \citep[e.g.][]{Bochsler2007,Laming2017}. The latter can be, however, measured less accurately compared to the photospheric analysis methods. In the seminal paper, \citet{Grevesse1998} presented the O abundance of $8.83 \pm 0.06$ dex, based on 1D LTE methods. The most recent estimate, derived by means of a detailed NLTE analysis with 3D radiation-hydrodynamics (RHD) simulations of solar convection, is $\logo = 8.69~\pm~0.03$ dex \citep[][based on the 777 nm lines]{Amarsi2018}, which compares well within the uncertainties with the  estimate by \citet[][based on the forbidden 630 nm line]{Caffau2015} ($\logo = 8.73 \pm 0.02 \pm 0.05$ dex). Neither of these estimates, however, are satisfactory for the SSM calculations \citep{Villante2020}, as the internal structure of the present-day solar model - the sound speed profile, the depth of the convective envelope - does not compare well with independent constraints on its structure obtained by means of helioseismology. Various scenarios have been put forward to explain the mismatch, ranging from the under-estimated opacity in the interior (\citealt{Bailey2015}, but see \citealt{Nagayama2019}) to energy transport by dark matter particles \citep{Vincent2015}. The problem has, so far, not been convincingly explained by any of these scenarios.

In this work, we present a new spectroscopic analysis of the solar photospheric O abundance. We are motivated by the availability of new atomic data and recent 3D magneto-hydrodynamic simulations (MHD) of near-surface convection, the chromosphere, and the photosphere \citep{Carlsson2016}. Our model atom of O relies on new data describing collisional excitation and charge transfer with H atoms, photo-ionisation, and collisions between \oi~ and free electrons. We also analyse the uncertainty in the oscillator strengths of the transitions observed. In addition, we develop a new model atom of Ni in order to study the influence of the Ni blend in one of the diagnostic \oi~features. Finally, we investigate the \oi~line formation using different types of the solar model atmospheres, including the \texttt{Stagger} RHD model, but also two 3D MHD models computed self-consistently, with and without chromosphere, using the \texttt{Bifrost} code. This is important for any study that requires precision abundance diagnostics, such as, e.g., a detailed analysis of the Sun relative to solar twins, comparative studies of exoplanet hosts, or spectroscopic age indicators \citep[e.g.][]{Bedell2014, Buchhave2015,Bedell2018, Lorenzo-Oliveira2018, Nissen2020}.

The paper is organised as follows. Section \ref{sec:observations} provides the details of observational datasets used in the abundance calculations. In Sect. \ref{sec:methods} we describe the main properties of the atomic models of O (Sect. \ref{sec:atomO}) and Ni (Sect. \ref{sec:atomNi}), in particular, the calculations of new photo-ionisation cross-sections and electron collision rates (Sect. \ref{sec:photo}), collisions with hydrogen atoms (Sect. \ref{sec:hcol}). In Sect. \ref{sec:atmos} we review the 1D and 3D (M)RHD model atmospheres, the statistical equilibrium and radiation transfer codes (Sect. \ref{sec:stateq}), and the details of abundance analysis (Sect. \ref{sec:abundcal}). We then describe the results of our calculations in Sect. \ref{sec:results}, addressing the following aspects: line formation of O (Sect. \ref{sec:linforO}) and Ni (Sect. \ref{sec:linforNi}), center-to-limb (CLV) variation (Sect. \ref{sec:finalCLV}), error analysis (Sect. \ref{sec:errors}), the solar O abundance (Sect. \ref{sec:likelihood}). We perform a comparative analysis of our findings in the context of earlier studies in the literature (Sect. \ref{sec:disc}). Finally Sect. \ref{sec:concl} summarizes the conclusions and provides an outlook for future work.
\section{Observations}\label{sec:observations}
In this work, we make use of different spectroscopic data available for the Sun.

Our primary dataset are the new spatially-resolved solar data obtained with the Fourier Transform Spectrograph mounted on the Vacuum Vertical Telescope at the Institut f\"ur Astrophysik G\"ottingen (hereafter, referred to as IAG data). Light from the 50cm-siderostat is picked up by a fiber with an on-sky diameter of 25" and sent to the FTS (for more information, see \citet{2016A&A...587A..65R} and \citet{10.1117/12.2560156}). The FTS was operated in double-sided mode with a spectral resolution of 0.024~cm$^{-1}$, i.e., $R = \frac{\Delta\lambda}{\lambda} \approx 700\,000$ at $\lambda=6\,000$ \AA). We collected $123$ individual observations of the quiet Sun at 12 different $\mu$-angles during April and August 2020. Each observation took approximately 10 minutes. We used the HITRAN database to identify telluric lines from H$_2$O and O$_2$ and mask them in the spectra. Relative motion between the VVT and the respective solar position was determined using the ephemeris code \citep{2015PhDT.......200D} based on the NASA's Navigation and Ancillary Information Facility SPICE toolkit \citep{1996P&SS...44...65A}. We used the differential rotation law of the Sun \citep{1990ApJ...351..309S} to correct the individual spectra for solar rotation, and we subtracted barycentric and rotational motion from every observed spectrum. Before addition, we normalized individual spectra and corrected remaining radial velocity offsets between different exposures taken at the same $\mu$-angles but at different positions on the Sun. We verified that these offsets were within the range of pointing errors. Eventually, we added spectra observed at the same $\mu$-angles. Our co-added spectra cover the spectral range between $4700$\AA and $7900$\AA and have a signal-to-noise (S/N) of 450 per resolution element at 6000\AA. In our analysis we employ the $\mu$ angles\footnote{Here $\mu = \cos \theta$ and $\theta$ is the heliocentric viewing angle, with the disc center corresponding to $\theta = 0^{\circ}$.}: $0.2$, $0.4$, $0.6$, $0.8$, $1.0$.

We also employ high-resolution spatially resolved spectra \citep{Pereira2009a, Pereira2009b} obtained with the TRIPPEL (TRI-Port Polarimetric Echelle-Littrow spectrograph) spectrograph mounted on the Swedish 1-m Solar Telescope (hereafter, referred to as SST data) \citep{Scharmer2003}. These data have a resolving power of $R = \lambda/\delta \lambda \approx 200\,000$ and cover the wavelength regions around the \oi~triplet (7771 to 7775 \AA) and [O~I] (6300 \AA) lines.

In addition, we analyse the spatially resolved solar spectra obtained with the Solar Optical Telescope on the Hinode satellite \citep{Caffau2015}. These data are not affected by telluric absorption, which is especially critical in the region around 630 nm. The HINODE data have a resolving power of $R = 200\,000$ and a S/N of $20\,000$. The pointings include the following $\mu$ angles: $0.38$, $0.70$, $0.86$, $0.95$, and $0.99$. For a more detailed description of the observations and the data reduction, we refer the reader to \citet{Caffau2015}. 

We complement the afore-mentioned data with the very high-quality,  $R \approx 350\,000$, disc-center spectra obtained using the KPNO FTS instrument. These data were released by \citet{Brault1972} and a re-reduced version was published in \citet{Neckel1984}. 
\begin{figure}
\includegraphics[width=0.45\textwidth, angle=0]{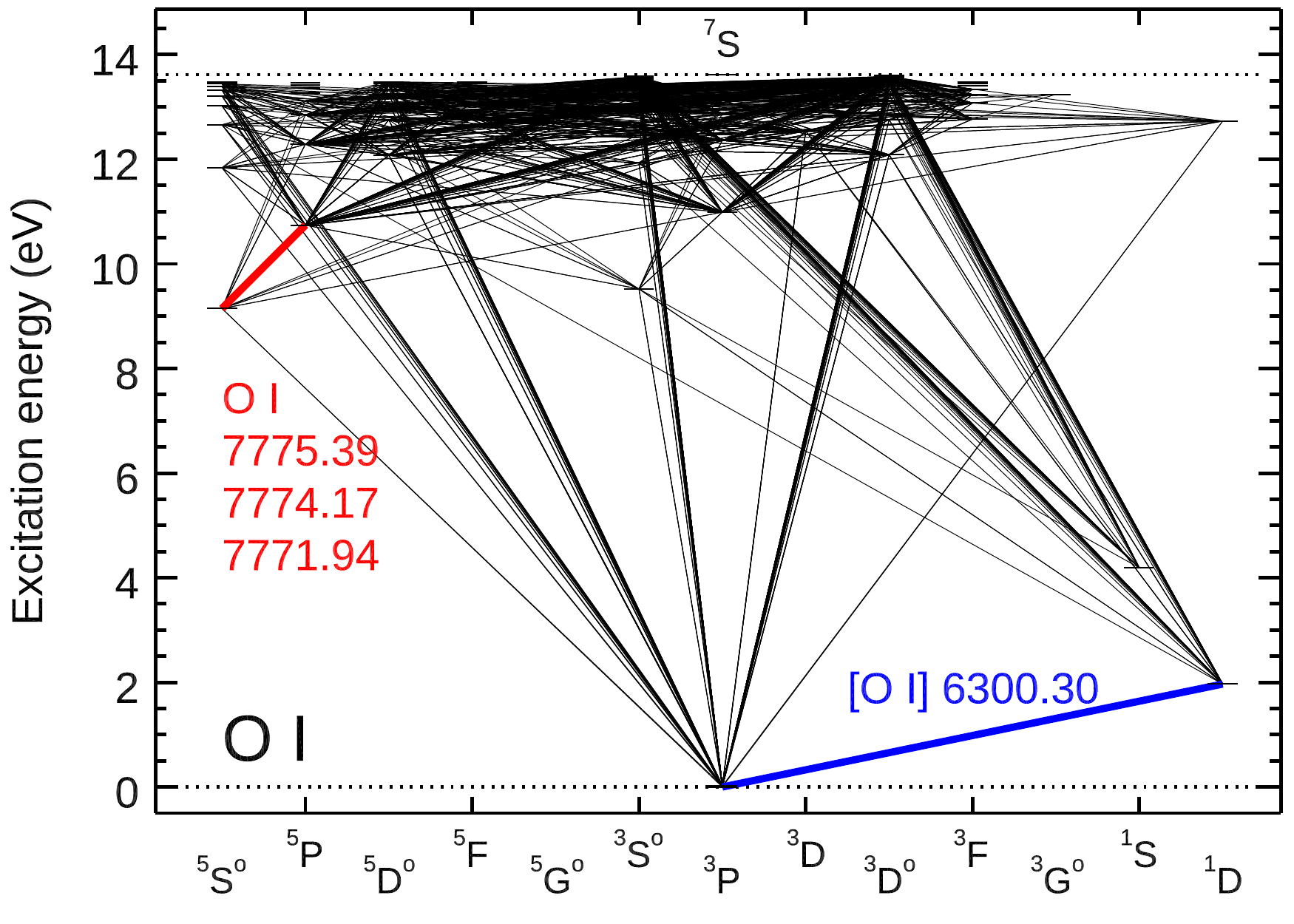}
\includegraphics[width=0.45\textwidth, angle=0]{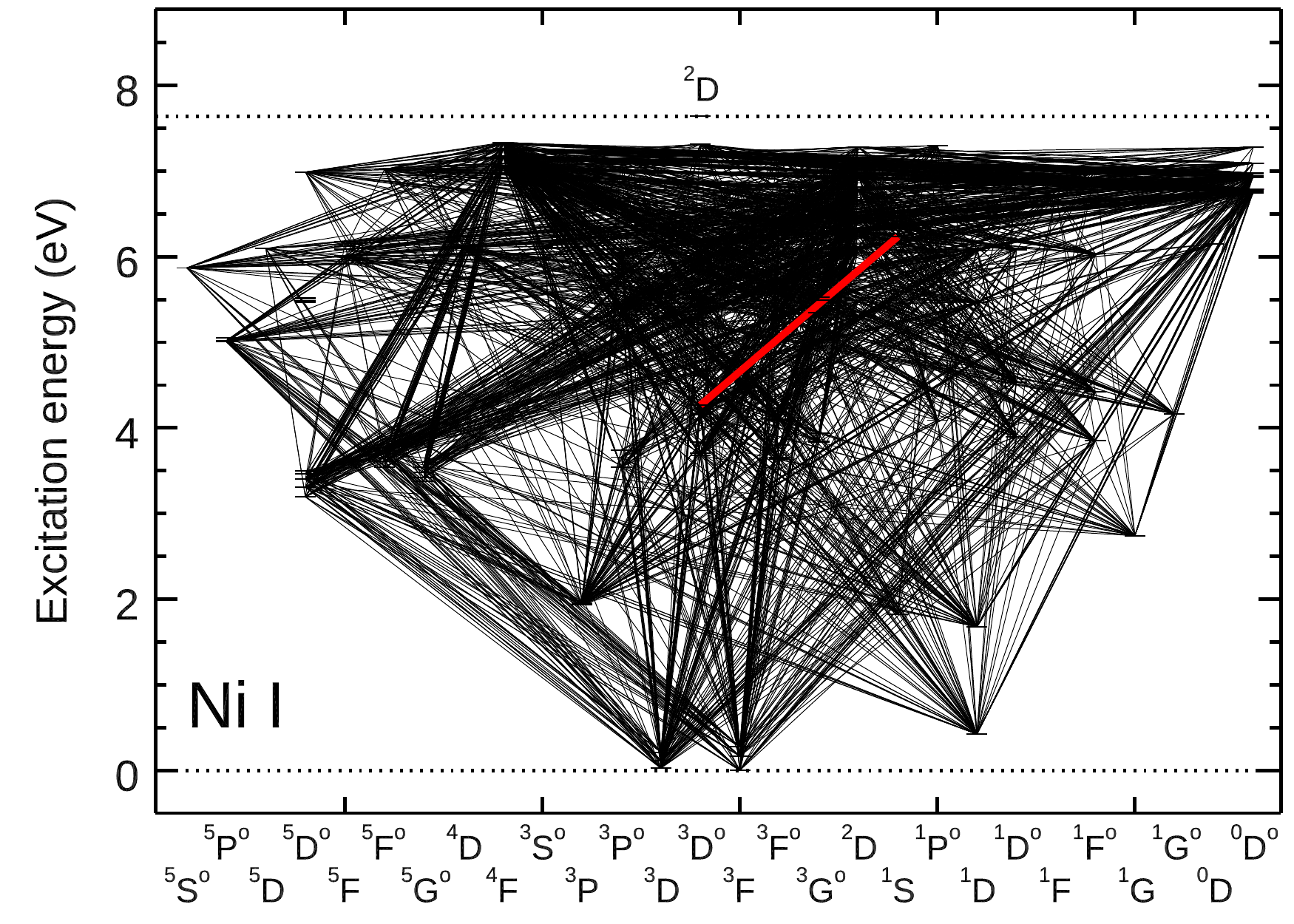}
\caption{Grotrian models of O (top panel) and Ni (bottom panel). The diagnostic features used in the spectroscopic analysis are indicated using bold blue and red lines.}
\label{fig:grotrian}
\end{figure}

The last resource of observational data in this work is the data collected by \citet{takeda2019} using the 60 cm Domeless Solar Telescope (hereafter, referred to as DST data) with the Horizontal Spectrograph at Hida Observatory of Kyoto University. In total, 31 different $\mu$-angles have been observed, of which we were only interested in the disc-center intensity. The spectra have a resolving power of $R \sim 140\,000$ and S/N of several hundreds.
\section{Methods}\label{sec:methods}

\subsection{O model atom}\label{sec:atomO}
The model atom consists of $122$ energy states, with neutral oxygen represented by 120 energy states and the singly-ionised oxygen by $2$ states, the ground state and the first excited state. The energy levels, which includes levels with principal quantum number $n\le 30$ and orbital angular momentum $l\le 4$, were assembled from the NIST database \citep{NIST_ASD}. The original number of \oi~ energy levels is $227$, however, we merge the fine structure levels with the energy above $102\,700$ cm$^{-1}$ (12.73 eV), so that all states above the level \Ot{3s}{1}{D}{o}{} are represented by a term with the statistical weight $g$ obtained by summing the individual statistical weights of the levels and the energy computed by weighted averaging of fine structure level energies according to their $g$-value \citep[cf.][]{Bergemann2012}. As a result our \oi~model contains $35$ fine structure states and $87$ terms. Fig. \ref{fig:grotrian} (top panel) shows the Grotrian diagram of \oi, with the key diagnostic transitions in the solar spectrum indicated with bold lines. The majority of lines in the \oi~ system connect energy levels with a very high , $E > 10$ eV, excitation potential (relative to the ground state). 

The diagnostic O lines used in the abundance analysis are the triplet \oi~lines at $777$ nm and the forbidden [O~I] line at 630 nm (Table \ref{tab:lines}). The former lines arise in electric dipole transitions from the 2p$^3$3s~$^5$S$^o_2$ level at $9.15$ eV above the ground level. The line at 630 nm is a magnetic dipole transition that arises from the ground level 2p$^4$~$^3$P$_2$. This transition is also observed in emission in the airglow spectra in the terrestrial atmosphere \citep{Reese1989, Slanger2011}.  Owing to its very low transition probability ($\log gf = -9.72$, \citealt{Storey2000}), the [O~I] line is weak even in the solar spectrum and its equivalent width (EW) measures only a few m\AA~(see Appendix). According to the NIST database\footnote{\url{https://physics.nist.gov/PhysRefData/ASD/Html/lineshelp.html}, accessed on and before May 22, 2021. Note the link only works in the copy-and-paste mode in a browser.}, the f-values of the 777 nm and the 630 nm transitions have an uncertainty rating "A" ($\leq 3\%$, or better than 0.01 dex) and "B+" ($\leq 7\%$, or better than $0.03$ dex), respectively. However, these uncertainty estimates from NIST appear to be highly overoptimistic. The true uncertainties are likely significantly larger, as indicated by the recently published new transition probabilities \citet{Civis2018} and our own calculations (Sect. \ref{sec:errors}). In this work, we have chosen to perform all abundance calculations using the average of \citet{Hibbert1991} and \citet{Civis2018} f-values for the 777 nm lines ($\log gf = 0.343, 0.197, -0.025$ for the 7771, 7774, and 7775 \AA~lines, respectively), and \citet{Storey2000} data for the 630~nm lines. We furthermore use one half of the difference between \citet{Hibbert1991} and \citet{Civis2018} ($0.026$ dex) as a $1\sigma$ uncertainty, and adopt a 20\% uncertainty ($0.08$ dex) on the f-value of the magnetic dipole line. We return to this issue in Sect. \ref{sec:errors}. The parameters of the diagnostic lines of \oi~are provided in Table \ref{tab:lines}.

The parameters of all other radiative bound-bound transitions in the O model atom, including wavelength, f-values, damping constants, were taken from the Kurucz \footnote{\url{http://kurucz.harvard.edu/atoms/0800/}} database. There are $13\,483$ radiative transitions in the wavelength range from $666$ \AA\ to $100\,000$ \AA.  However, after merging the energy states and making a cut on the oscillator strength ($\log gf = -10$), the number of radiative transitions is compressed to $1364$. The chosen cut is needed to capture the resolved photoionisation and recombination resonances, including the Rydberg Enhanced Recombination \citep{Nemer2019}. These processes lead to a sequence of transitions that involve the diagnostic O I lines, hence they are necessary for the accuracy of the SE calculations. The lines long-ward of $20\,000$ \Angstrem~ are important, because they represent a channel through which the majority of very high-excitation states are connected and therefore, ensure their convergence to statistical equilibrium (SE). 

Most of the lines in the atomic model are represented by Voigt profiles with $9$ frequency points, which is sufficient for the SE calculations. The diagnostic \oi~lines are represented by $251$ frequency points. Van der Waals damping, that is, broadening caused by elastic collisions with H atoms, is included using the data from \citet{Barklem2000}. It should be noted, though that only $33$ of our lines are present in the \citet{Barklem2000} database, therefore for the remainder of transitions we resort to the standard Uns{\"o}ld formalism \citep{Unsold1955}. It is known that the Uns{\"o}ld theory underestimates line broadening, as it only applies to collisional broadening at large atomic separations. However, this is not of a concern for \oi, as our tests with Uns{\"o}ld damping constants scaled by a factor or $2$ and $0.5$ yield virtually identical results for the SE and the profiles of diagnostic lines. 

\begin{table*}
\begin{minipage}{\linewidth}
\renewcommand{\footnoterule}{} 
\setlength{\tabcolsep}{2.5pt}
\caption{Parameters of spectral lines used for the solar O abundance calculation. The wavelengths for the diagnostic lines are taken from the NIST database. References to the $\log gf$ values and H damping are also provided.}
\label{tab:lines}     
\begin{center}
\begin{tabular}{l c cc cc cc cc}
\noalign{\smallskip}\hline\noalign{\smallskip}  Specie & $\lambda$ & $\Elow$ & $\Eup$ & Lower & Upper & $\log gf$ & vdW & Ref ($\log gf$) & Ref (vdW) \\
  &  ~~~~[\AA] & [eV] & [eV] & level & level & & & & \\
\noalign{\smallskip}\hline\noalign{\smallskip}
O & & & & & & & & & \\
O I & 7771.940  & 9.146 & 10.741 & \Ot{3s}{5}{S}{o}{2}   & \Ot{3p}{5}{P}{}{3}    & $0.369$, $0.317$ & 453.234 & \citet{Hibbert1991}, \citet{Civis2018}  & \citet{Barklem2000} \\
O I & 7774.170  & 9.146 & 10.741 & \Ot{3s}{5}{S}{o}{2}   & \Ot{3p}{5}{P}{}{2}    & $0.223$, $0.170$ & 453.234 & \citet{Hibbert1991}, \citet{Civis2018}  & \citet{Barklem2000} \\
O I & 7775.390  & 9.146 & 10.740 & \Ot{3s}{5}{S}{o}{2}   & \Ot{3p}{5}{P}{}{1}    & $0.002$, $-0.051$ & 453.234 & \citet{Hibbert1991}, \citet{Civis2018}  & \citet{Barklem2000} \\
O I & 6300.304  & 0.000 &  1.967 & \Ot{2p^4}{3}{P}{ }{2} & \Ot{2p^4}{1}{D}{}{2}  &  -9.72 &  & \citet{Storey2000} & \citet{Unsold1955} \\
Ni & & & & & & \\
Ni I  &  6300.341  &  4.266  & 6.234 & \Nt{4p}{3}{D}{o}{1} & \Nt{4s^2}{1}{S}{}{0} & -2.11 &  & \citet{Johansson2003} &  \citet{Unsold1955} \\
\noalign{\smallskip}\hline\noalign{\smallskip}
\end{tabular}
\end{center}
\end{minipage}
\end{table*}

\subsection{Photoionisation and electron collision rates}\label{sec:photo}

The standard source of photoionisation cross-sections is the Opacity Project (OP) database  TOPbase\footnote{\url{http://cdsweb.u-strasbg.fr/topbase/topbase.html}} \citep{Cunto1992}. However, the data were computed at a very coarse energy resolution assuming the $LS$ coupling and they are available only for the energy states with the principal quantum number of $n\le 10$. These shortcomings limit the applicability of the OP data to NLTE stellar atmosphere problems.

Also the data that describe the rates of transitions caused by inelastic collisions with free electrons are very sparse. The most recent source of this information is \citet{Barklem2007}, who tabulated the rate coefficients for transitions between the $7$ low-energy states in \oi. They employed the standard R-matrix method assuming LS coupling, which is known to deliver relatively accurate results consistent with direct experimental values. These sparse datasets are, however, not sufficient to describe the entire micro-physics of interactions between \oi~atoms, radiation field, and electrons. This is especially important, because the key diagnostic \oi~ lines (the 777 nm triplet) connect very high-excitation energy levels (Fig. \ref{fig:grotrian}). Therefore, we opted to compute new photoionisation cross-sections and cross-sections for the transitions caused by e$-$ collisions. 

We computed new photoionisation cross-sections for all levels with a  principal quantum number up to $n=20$ (energies up to 109561 cm$^{-1}$). These calculations were carried out with the Breit-Pauli version of the code RMATRX \citep{Berrington1995}. This is an implementation of the R-matrix method for atomic scattering calculations. The atomic orbitals for the O$^{+}$ target system were obtained with the scaled Thomas-Fermi-Dirac-Amaldi central-field potential as implemented in {\sc AUTOSTRUCTURE} \citep{Badnell1997}. Our target representation included eleven configurations $2s^23p^3$, $2s^23p^23s$, $2s^23p^23p$, $2s^23p^23d$, $2s3p^4$, $2s3p^33s$, $2s3p^33p$, $2s3p^33d$, $2s^23p3s^2$, $2s^23p3p^2$, and $2s^23p3d^2$. This expansion lets to $26$ close coupling LS terms and $39$ energy levels with principal quantum number up to $n=20$. 
 
\begin{figure}
\includegraphics[width=0.48\textwidth, angle=0]{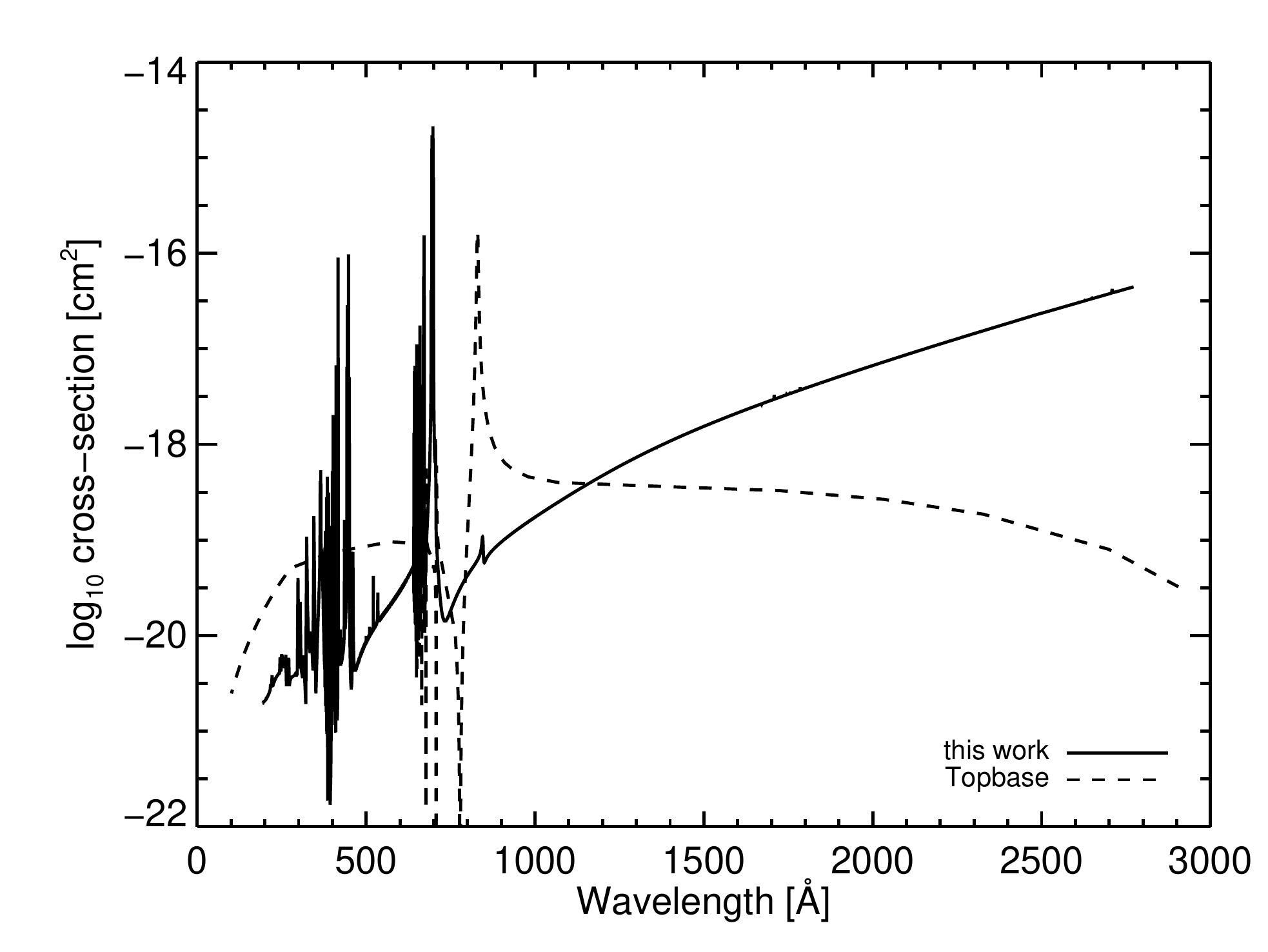}
\includegraphics[width=0.48\textwidth, angle=0]{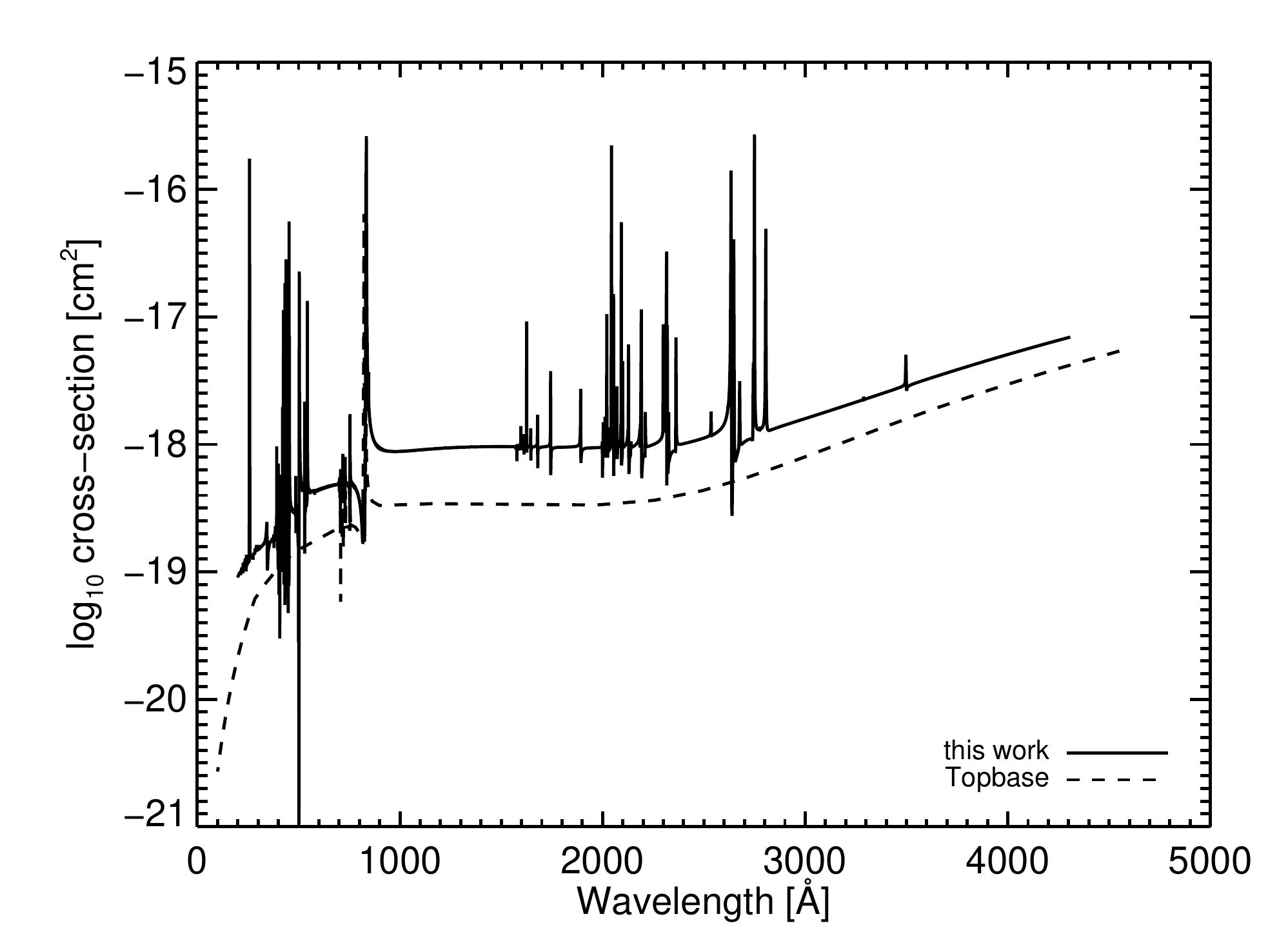}
\caption{New photoionisation cross-sections for the \Ot{3s}{5}{S}{o}{2} and \Ot{3p}{5}{P}{}{1} levels of \oi~compared with the Opacity Project data (here we take the cross-section in OP LS-coupling of the $^5$P for the $^5$P$_1$ level).}
\label{fig:photo}
\end{figure}

We give a preference to the new photoionisation cross-sections over the TOPbase values. First, our target and close coupling expansions are larger than those of the OP. Second, the present cross-sections are in intermediate coupling and account for relativistic corrections, as opposed to the LS-term OP cross sections. Finally, we retained level-to-level partial cross-sections, which allow for a reliable determination of recombination rates. Also, our cross sections have the energy resolution of $2.2\times 10^{-4}$~Ryd, (about $100\times$ better resolution than OP cross sections), which adequately resolves the cross-section resonances. Fig.~\ref{fig:photo} compares our new values for two selected energy levels with the TOPbase data. Our cross-sections include more resonance structures, owing to the much larger close-coupling expansion. In the case of the 3s~$^5$S$^o$ state, the TOPbase cross-section exhibits an abnormal decreasing trend at higher energies and a sort of bulge at lower energies. Our data show a correct high energy behaviour and reveal that the bulge is, in fact, a series of resonances, which only show up as proper close-coupling channels are included in the calculations. Regarding the cross-section for the 3p~$^5$P$_1$ level, the results in TOPbase and our data show very similar qualitative behavior, but our cross-section accounts for more series of resonances. It should be noted that TOPbase only provides data in $LS$ coupling, thus the cross-section show in Fig.~\ref{fig:photo} is actually the state's cross-section.

We also calculated new electron impact excitation rate coefficients using the AUTOSTRUCTURE code \citep{Badnell2011}, which is based on the Breit Pauli Distorted Wave method. All configurations of the form $2s^22p^4$, $2s^22p^3nl$, $2s2p^5$, and $2s2p^4nl$, with $n\le 30$ and $0\le l\le 3$ were included in the calculations. For the rest of the \oi~system, we complemented the R-matrix collision strengths with the rate coefficient computed using the formulae from \citet{Regemorter1962}.
\subsection{Collisions with hydrogen atoms}\label{sec:hcol}
\begin{table}
\begin{minipage}{\linewidth}
\renewcommand{\footnoterule}{} 
\caption{Scattering channels and the corresponding asymptotic energies, taken from NIST \citep{NIST_ASD}.}
\label{tab:states_OH}
\begin{center}
\begin{tabular}{clr}
\hline 
j & Scattering channels & Asymptotic energies \\
  &    &  eV     \\
\hline 
\hline 
1  & O$(2p^{4}\,^{3}$P) + H$(1s)$  & 0.0000     \\
2  & O$(2p^{4}\,^{1}$D) + H$(1s)$ & 1.9674  \\
3  & O$(2p^{4}\,^{1}$S) + H$(1s)$ & 4.1897  \\
4  & O$(2p^{3}3s\,^{5}$S$^{\circ}$) + H$(1s)$  & 9.1461  \\
5  & O$(2p^{3}3s\,^{3}$S$^{\circ}$) + H$(1s)$ & 9.5214  \\
6  & O$(2p^{4}\,^{3}$P) + H$(2s)$  & 10.2000 \\
7  & O$(2p^{4}\,^{3}$P) + H$(2p)$ & 10.2000 \\
8  & O$(2p^{3}3p\,^{5}$P) + H$(1s)$ & 10.7406 \\
9  & O$(2p^{3}3p\,^{3}$P) + H$(1s)$  & 10.9888 \\
10 & O$(2p^{3}4s\,^{5}$S$^{\circ}$) + H$(1s)$   & 11.8376 \\
11 & O$(2p^{3}4s\,^{3}$S$^{\circ}$) + H$(1s)$  & 11.9304 \\
12 & O$(2p^{3}3d\,^{5}$D$^{\circ}$) + H$(1s)$   & 12.0786 \\
13 & O$(2p^{3}3d\,^{3}$D$^{\circ}$) + H$(1s)$   & 12.0870 \\
14 & O$(2p^{3}4p\,^{5}$P) + H$(1s)$   & 12.2861 \\
15 & O$(2p^{3}4p\,^{3}$P) + H$(1s)$   & 12.3589 \\
16 & O$(2p^{3}3s\,^{3}$D$^{\circ}$)+H$(1s)$   & 12.5402\\
 & & \\
1 & O$^{-}(2p^{5}\,^{2}$P) + H$^{+}$ & 12.1500 \\
2 & O$^{+}(2p^{3}\,^{4}$S$^{\circ}$) + H$^{-}$($^{1}$S)  & 12.8641 \\
\hline 
\hline 
\end{tabular}
\end{center}
\end{minipage}
\end{table}

Inelastic collisions with H atoms represent an important ingredient in NLTE calculations, as they lead to excitation, de-excitation, ion-pair formation, and mutual neutralization reactions. Detailed quantum-mechanical calculations of the rate coefficients for O were recently presented by two independent groups, \citet{Barklem2018} and \citet{Belyaev2019}. These studies rely on different approaches. \citet{Barklem2018} estimated the long-range electronic structure of the OH molecule, by approximating the molecular wave function by the linear combination of atomic orbitals (LCAO) of the atoms that form the molecule, while the calculations of collisional dynamics were performed in the framework of the multichannel model \citep{Belyaev1993}, we refer to this approach as LCAO multichannel study. The O$^-$ + H$^+$ scattering channel was not taken into account. Independently, \citet{Mitrushchenkov2019} calculated the electronic structure of the OH molecule using the multi-reference configuration interaction (MRCI) method and estimated the rate coefficients for O + H, O$^+$ + H$^-$ and O$^-$ + H$^+$ collisional processes using the multi-channel model. Later on, \citet{Belyaev2019}, based on the MRCI electronic structure by \citet{Mitrushchenkov2019}, calculated the rate coefficients for inelastic processes in the same transitions using the quantum hopping probability current (QPC) method. The MRCI QPC study gives a more accurate description of inelastic collisions than the LCAO multichannel study, because it uses a more accurate electronic structure and take into account both the long- and short-range non-adiabatic regions, see \citet{Belyaev2019} for the detailed comparison. 

\citet{Barklem2018} tabulate the rate coefficients for the transitions between the lowest $18$ terms in \oi~(up to \Ot{3s}{1}{D}{o}{} at 102584 cm$^{-1}$), whereas \citet{Belyaev2019} provide the data for the lowest $11$ terms (up to \Ot{4s}{3}{S}{o}{} at 96147 cm$^{-1}$). In this work, we complement the latter data with new calculations for $5$ additional states of \oi~with excitation energies from $12.0786$ eV to $12.5402$ eV, as well as two ionic terms O$^+$ + H$^-$  and O$^-$ + H$^+$. All energy states included in the dynamical calculations are provided in Table~\ref{tab:states_OH}. For these $5$ additional states the transition probabilities are computed using the asymptotic approach \citep{Belyaev:2013pra}, including the non-adiabatic nuclear dynamical treatment accomplished by the multi-channel approach. The calculations are performed within the $^4\Sigma^-$ molecular symmetry. Higher-lying covalent states are not included in the calculations, because they either create ionic-covalent avoided crossings at inter-nuclear distances greater than $\sim 100$ atomic units, and hence the corresponding rates are negligible, or they do not create avoided crossings at all. It should be noted that non-adiabatic transitions between the higher-lying states are possible at short-range distances, although short-range transitions do not lead to high rate coefficients. These rates can be roughly estimated by means of the Kaulakys analytical approach \citep{Kaulakys1985, Kaulakys1991}. These datasets were recently computed by P. Barklem\footnote{\url{https://github.com/barklem/kaulakys}}, however, only for the lowest $18$ energy states of \oi~ that overlap with the states included in the LCAO multichannel calculations. None of the available datasets accounts for the fine structure, therefore, for the lack of a more accurate formulation, we assign the same rate coefficient to each fine structure level within a given term. We note that an alternative approach was proposed in the literature \citep[e.g.][]{Osorio2015}, however, also this approach involves an assumption of the relative probability of transitions to different final spin states. Our extensive tests suggest that reducing or increasing the quantum-mechanical rate coefficients by a factor of $3$ (which corresponds to the average multiplicity of \oi) does not have any significant influence on the SE of oxygen, in line with the findings of \citet{Bergemann2019}, thus any further arbitrary manipulation of the datasets is not justified. 
\begin{figure}
%\hbox{
\includegraphics[width=0.47\textwidth,angle=0]{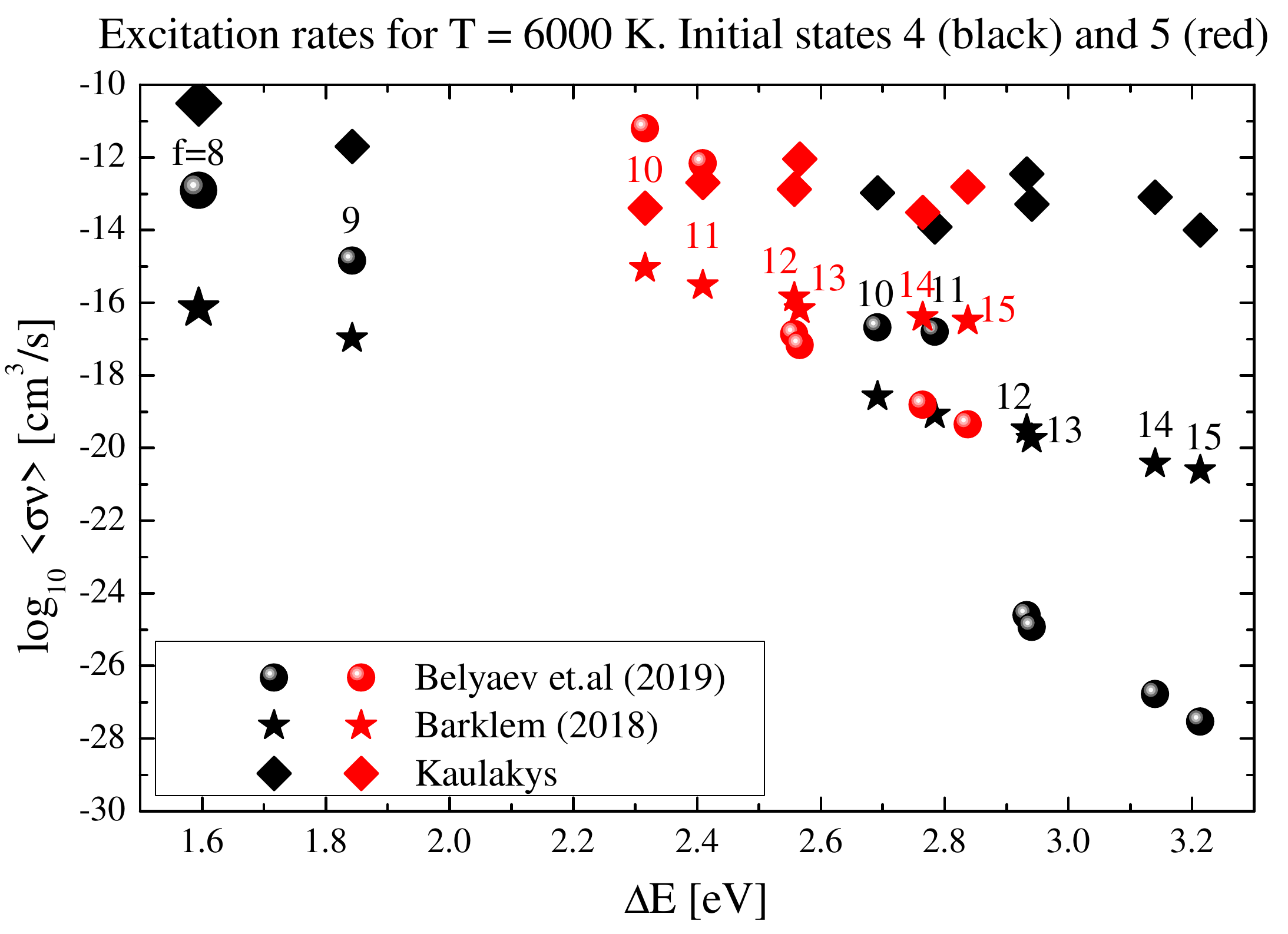}
\includegraphics[width=0.47\textwidth,angle=0]{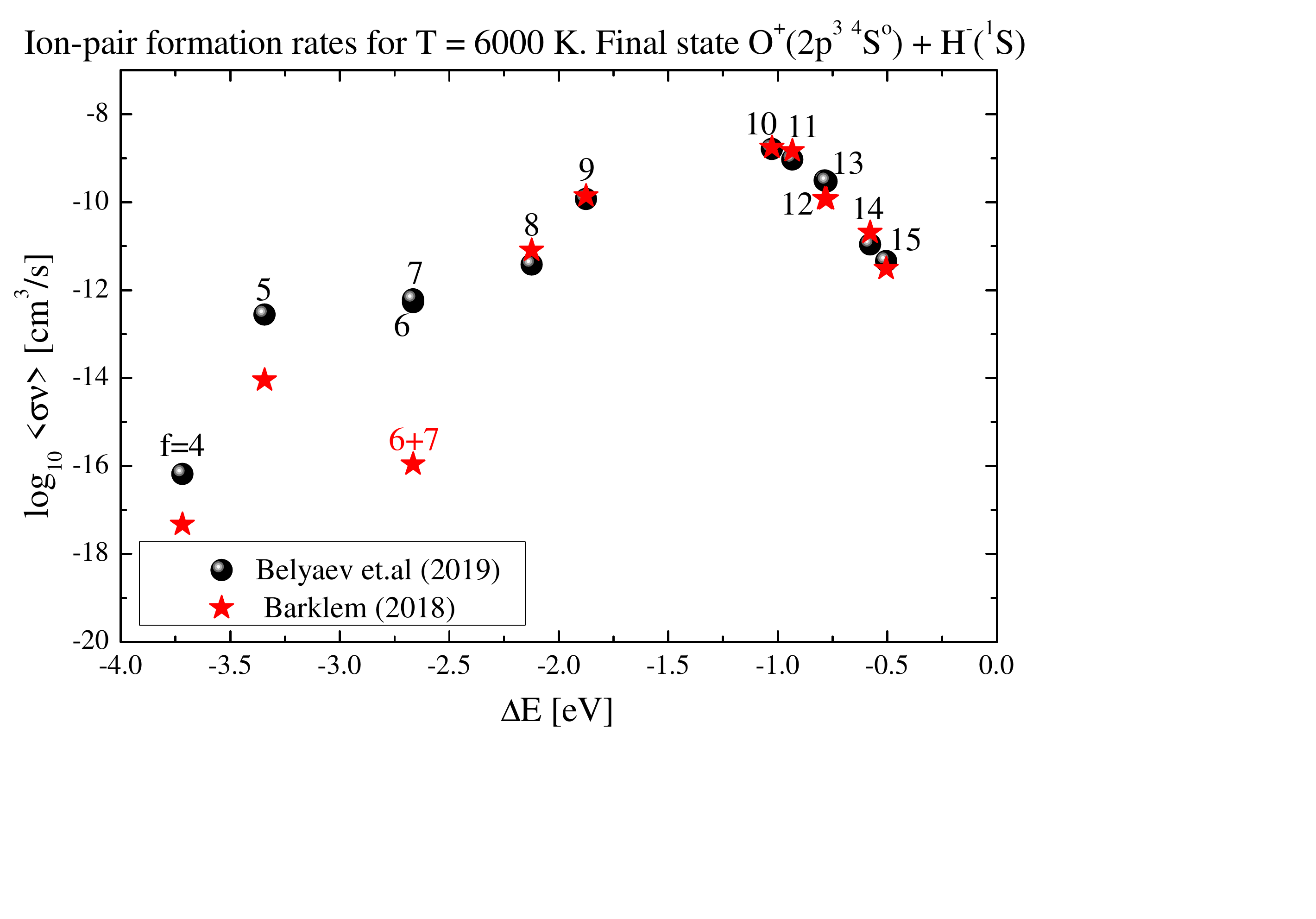}
%}
\caption{Rate coefficients of the excitation ($\Delta E > 0$) and ion-pair formation processes in O $+$ H collisions at  $T = 6\,000$ K as a function of the energy defect $\Delta E$ for the initial (top panel) and final (bottom panel) channels $i = 4$ and $i = 5$ (Table 2). Three sets of data, \citet{Barklem2018}, \citet{Belyaev2019}, and \citet{Kaulakys1991}, are shown.}
\label{fig:collisions}
\end{figure}

Fig. \ref{fig:collisions} (top panel) shows the data for some the transitions in common between \citet{Barklem2018}, our new estimates, and the values estimated using the model of \citet{Kaulakys1985,Kaulakys1991}. Interestingly, there is no systematic difference between the two approaches. For some transitions, \citet{Barklem2018} provide significantly, by up to $4$ orders of magnitude, lower rate coefficients. The difference can be explained by the fact that the MRCI QPC study takes non-adiabatic regions at small inter-nuclear distances into account. Due to the transitions at a short range, the probability currents are redistributed differently, as compared to the simplified picture given by including long range non-adiabatic regions (as done in the LCAO multichannel study) only. This also explain the difference in the rates for the important transitions, \Ot{3s}{5}{S}{}{o} - \Ot{3s}{3}{S}{}{o} and \Ot{3s}{5}{S}{}{o} - \Ot{3p}{5}{P}{}{}. The rates computed using the \citet{Kaulakys1985,Kaulakys1991} recipe are typically much larger compared to the data computed using the LCAO multichannel or the MRCI QPC studies. The differences are most significant for transitions with $\Delta E \gtrsim 2.5$ eV. In this regime, the Kaulakys model predicts the rate coefficients that are up to $10$ orders of magnitude higher compared to the detailed quantum-mechanical calculations. This is not surprising, as the model was developed to describe transitions between high-energy Rydberg states. 

Fig. \ref{fig:collisions} (bottom panel) illustrates the rate coefficients of ion-pair formation processes computed using the MRCI QPC and the LCAO multichannel methods. There is no significant difference for the majority of the ion-pair formation processes, except the processes with $\Delta E < -2.5$ eV. But the general behaviour of all ion-pair formation processes is very similar.

It is worth pointing out that the scattering channels $j=9 - 13$ belong to the so-called 'optimal window', in accordance with the general behavior found by applying the simplified model \citep{BelyaevYakovleva2017aa}. The model predicts the largest rates for mutual neutralization processes with final-state binding energies in the vicinity of $-2$ eV, that corresponds to the excitation energy of $11.65$ eV for the OH molecule.  This optimal window is well understood through the simplified model and it describes the mechanisms of inelastic atomic collision processes due to long-range non-adiabatic regions created by the ionic-covalent interaction. 

Since the LCAO multichannel, MRCI QPC, and Kaulakys data represent physically different approaches to the determination of H impact transition rates, we proceed as follows. We create three NLTE atomic models of O and use them independently in the statistical equilibrium and radiative transfer calculations. One of the models includes the LCAO multichannel data only (hereafter, the LCAO model atom). The other model relies on the LCAO multichannel data co-added with Kaulakys rate coefficients (hereafter, the LCAO$+$Kaul model), which is appropriate because the LCAO multichannel data describe long-range interactions only and the Kaulakys data are expected to compensate for the lack of short-range interactions. The third model (hereafter, the QPC model) relies on the new rates computed using the MRCI molecular structure and the hopping probability current method. Since in all atomic models, less than $20$ energy states (out of $120$ \oi~states in the model atom) are represented with detailed quantum-mechanical data, we connect all other energy states by collision-induced transitions with the rate coefficients computed using the scaled Drawin's formula \citep{Drawin1968}. As we show in Sect. \ref{sec:nltecor}, the only effect of this approach is that it ensures the collisional coupling with the entire \oi~ system and, therefore, helps to greatly speed up the convergence in the very time-consuming 3D NLTE calculations. Both atomic models also include the rates of charge transfer between the O$^+$ and H$^-$ ions, computed self-consistently using the methods discussed above.
\subsection{Ni model atom}\label{sec:atomNi}
Since one of the critical diagnostic \oi~ lines is blended by Ni, we have opted for performing a detailed study of NLTE effects in Ni. To the best of our knowledge, this is the first analysis of the  statistical equilibrium of Ni with a comprehensive atomic model with 1D and 3D model atmospheres.

Similar to O, the atomic data for \nii~ and \niii~ were adopted from the Kurucz database\footnote{Taken from \url{http://kurucz.harvard.edu/atoms/2800/}, using the files alitzen2800.dat,  b2800e.com,  and b2800o.com}. The initial datasets comprise $281$ energy levels and $9663$ radiative transitions for \nii, and $716$ energy levels and $56\,193$ radiative transitions for \niii, respectively. The \nii~levels are taken from the \citet{Litzen1993} study, which combines experimental measurements with theoretical calculations of the energy level structure. We merge fine structure levels below $6.237$ eV and remove transitions with $\log gf < -10$. The uppermost Ni I level in our model has the energy of $7.34$ eV, whereas the first ionisation threshold (ground state of Ni II) is located at $7.64$ eV. We note that there are only two known energy levels of Ni I that are higher than our threshold for Ni I and are below the 1st ionisation threshold. One of them has the energy of 7.43 eV (59940.517 cm$^{-1}$) and it represents one of the fine structure levels of the 3d8(3F)4s(4F)5p $^5$D$^o$ term. This level is, therefore, merged with the other lower-energy FS levels of this term. The other level is the 3d8(3F)4s(4F)6s $^5$F state with the energy of 7.43 eV (59862.611 cm$^{-1}$). Owing to its small J-value ($J = 5$) and a very small number of lines connecting it with the rest of Ni I, we merge it with the Ni I state at  7.34 eV. We have verified that our NLTE results do not depend on the treatment of these 2 levels. Our final model of Ni comprises $538$ energy states and $2468$ spectral lines represented by Voigt profiles with a $9$ point frequency quadrature. The first ionisation threshold of \nii~is located at $7.64$ eV and that of \niii~ at $18.17$ eV. The Grotrian diagram of \nii~is shown in Fig. \ref{fig:grotrian} (bottom panel). Ni has a very complex energy level system with many singlet, triplet, and quintet terms. It should be noted that this diagram only shows energy levels with an LS designation and it is not complete, as many energy levels in \nii~ are described in J$_{c}$K coupling scheme \citep{Litzen1993}.

Our model atom does not include isotopic splitting and hyperfine structure (for the odd Ni components), as the shifts are very small. Using the data from \citet{Johansson2003}, we find, for example, that including the two main isotopes of Ni changes the equivalent width of the \nii~ line at $6300.342$ \Angstrem~ by less than $0.5 \%$. Therefore, we neglect isotopic shift in the SE calculations of Ni and in the abundance analysis of O. 

The photoionisation cross-sections were computed using the hydrogenic approximation. The rates of collisional excitation by electrons and hydrogen atoms were calculated using the standard formulae of \citet{Regemorter1962} and \citet{Drawin1968}. We apply a scaling factor of $0.05$ to the rates that describe the collisions of \nii~atoms with H, because the standard formulae are known to severely over-estimate the rate coefficients compared to the detailed quantum-mechanical data \citep{Barklem2016}. However, we investigate in Sect. \ref{sec:results} how this assumption influences the results. The rates of collisional ionisation were computed using the \citet{Seaton1962} formula. 
\subsection{Model atmospheres}\label{sec:atmos}
\begin{figure}
\includegraphics[scale=0.7]{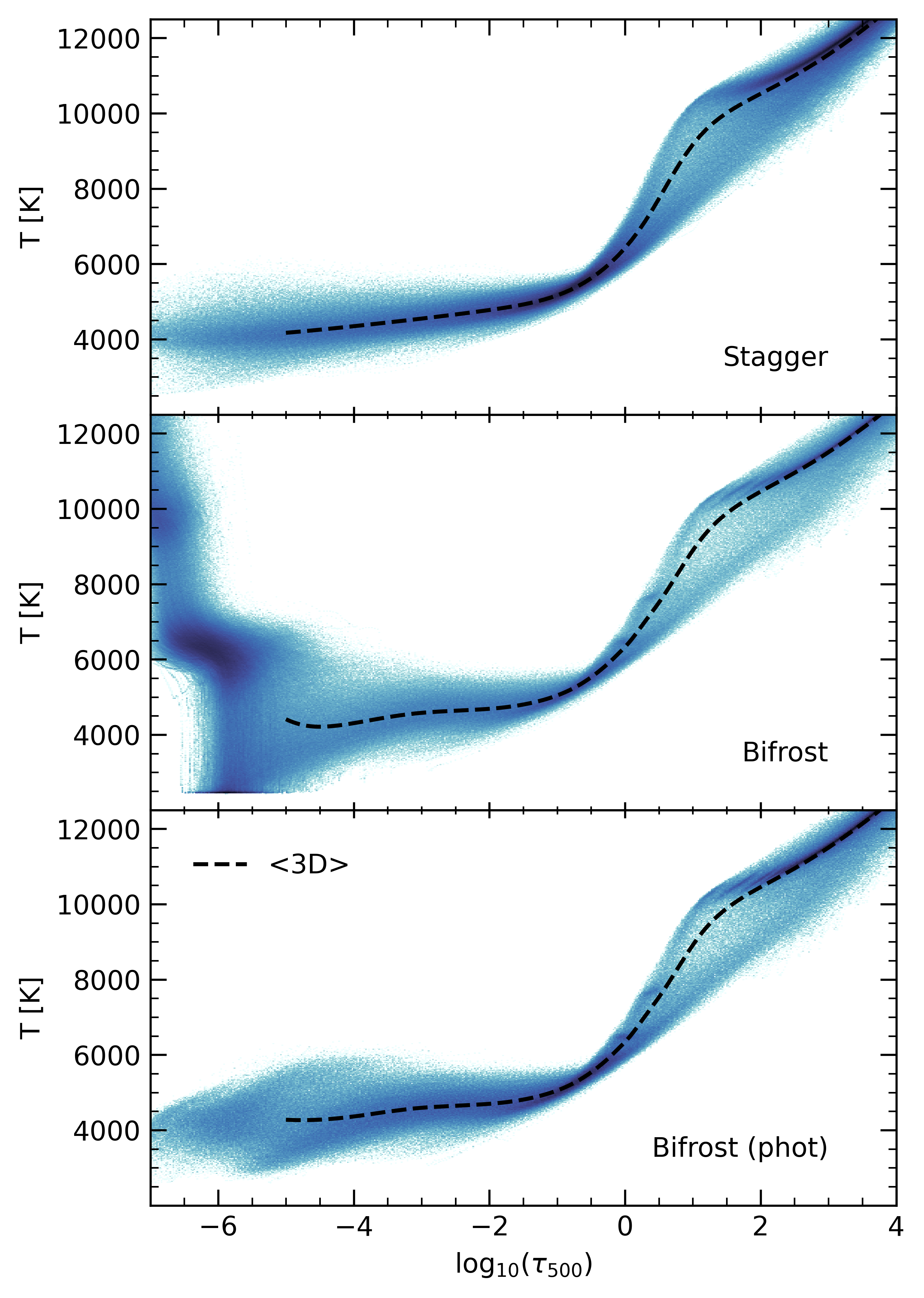} 
\caption{Temperature structure of the \texttt{STAGGER} and \texttt{Bifrost} 3D (M)RHD models as a function of logarithmic optical depth at 500 nm $\opd$. The dashed lines show the T($\tau$) structure of the corresponding \tda\ averages. See text}.
\label{fig:STAPH}
\end{figure}

\begin{figure*}
\centering
\hbox{
\includegraphics[scale=0.6]{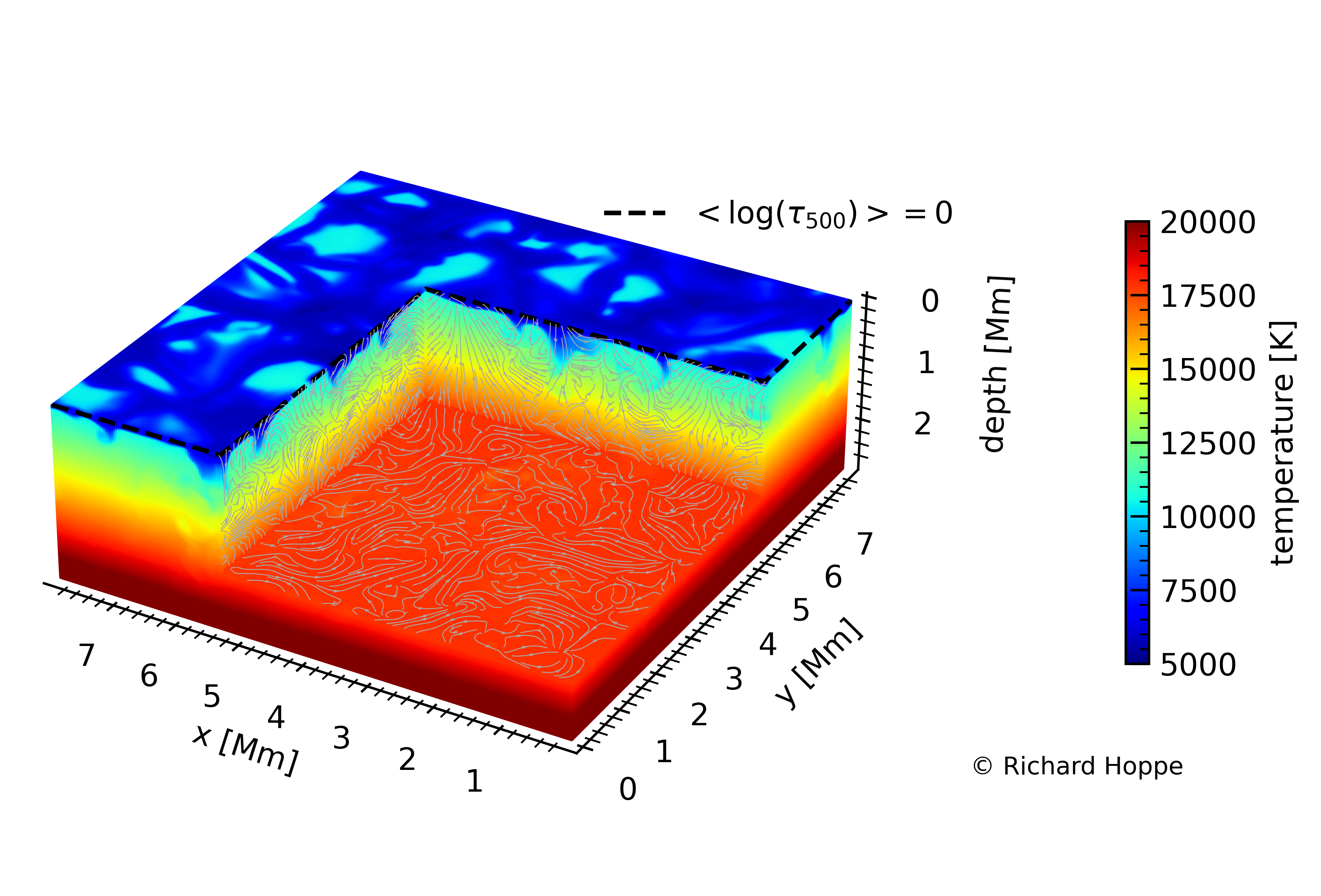}
\includegraphics[scale=0.6]{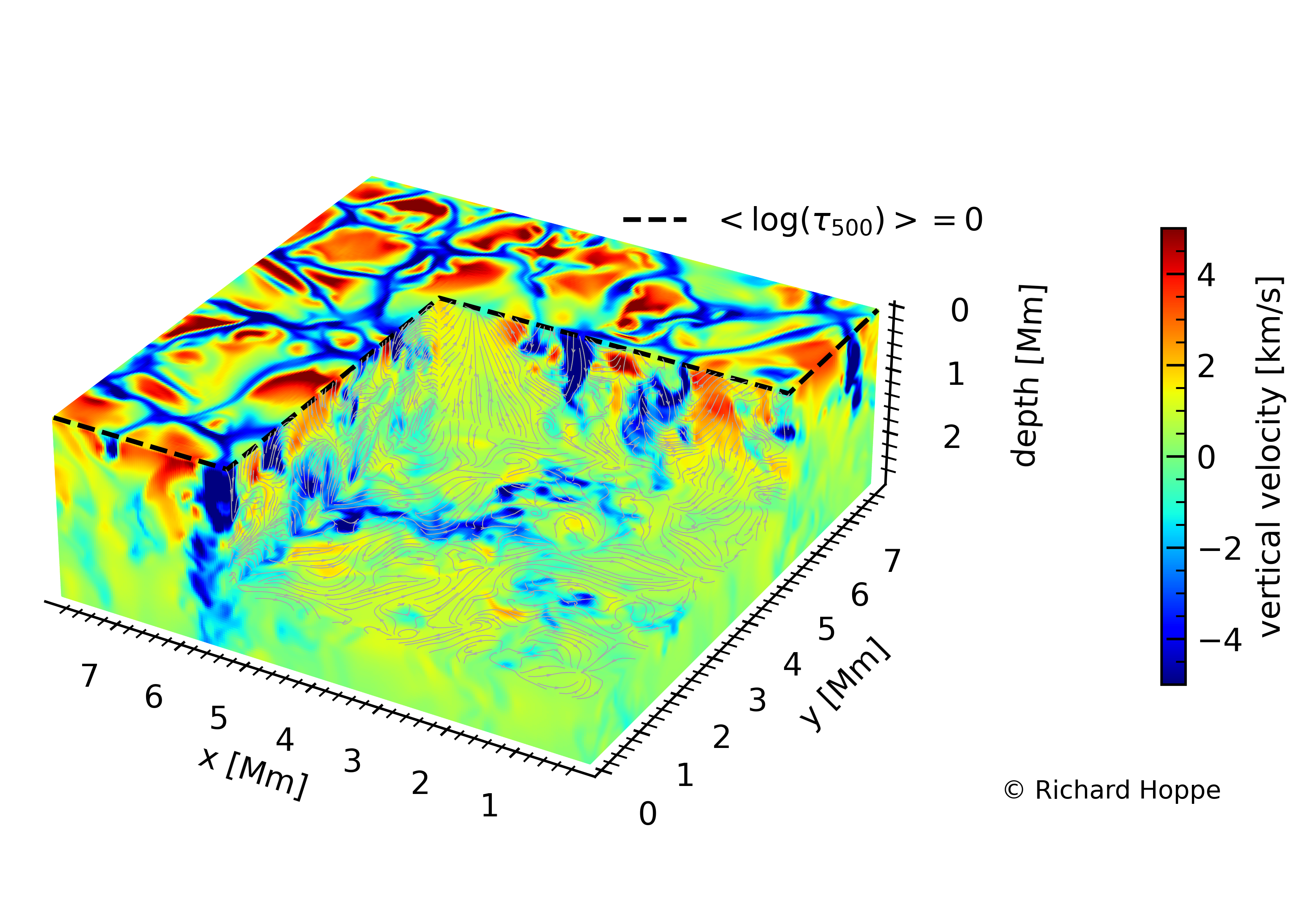}
}
\hbox{
\includegraphics[scale=0.6]{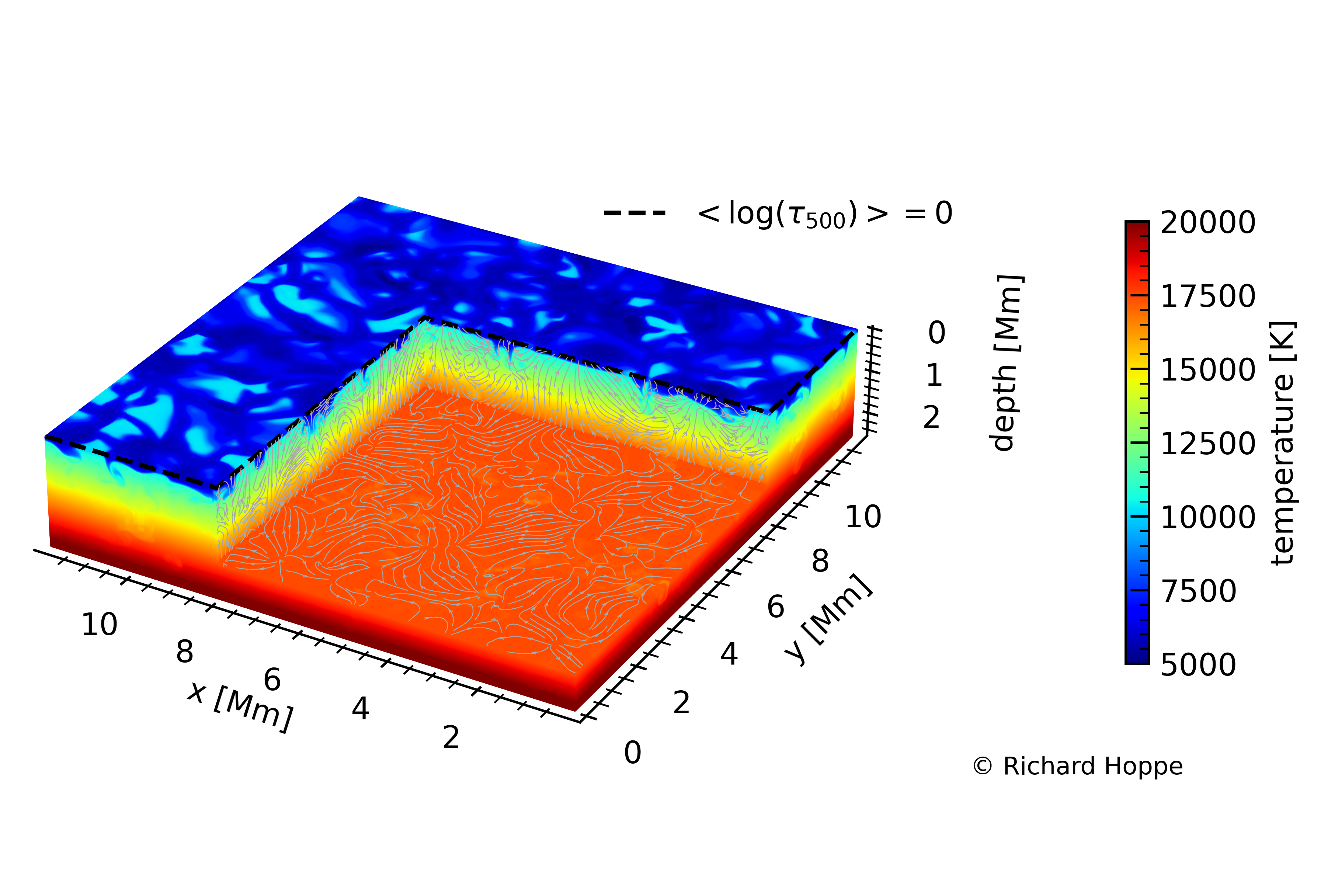}
\includegraphics[scale=0.6]{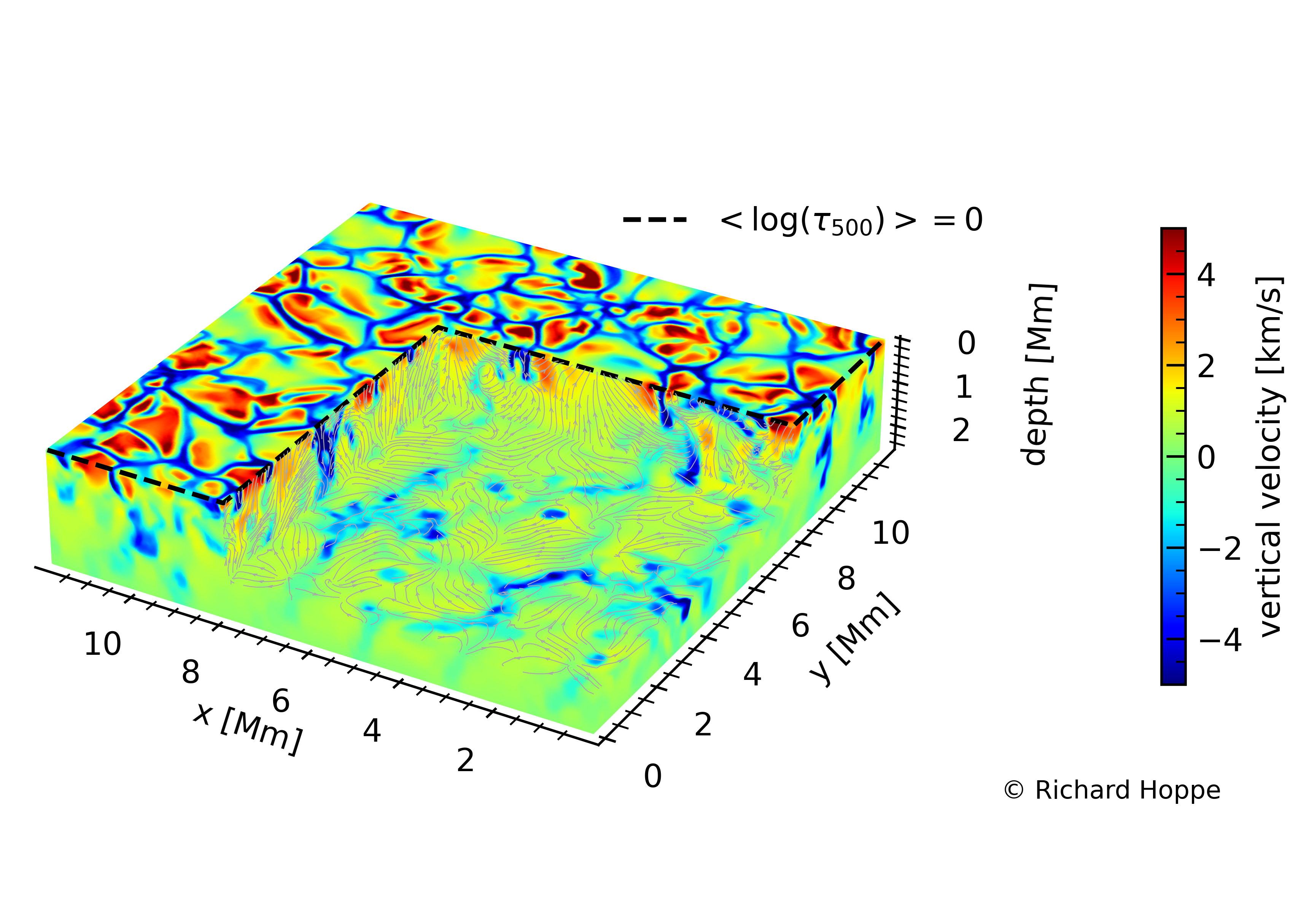}
}
\caption{Temperature and vertical velocity distributions of a representative Stagger (top) and Bifrost (bottom) model atmosphere snapshots. Both models have the full horizontal resolution, but, for illustration, we do not show the part of the atmosphere above $\opd \approx 0$. The location of the horizontally averaged optical depth surface is indicated with a dashed line.}
\label{fig:StaggerTemp}
\end{figure*}
We use several types of the solar model atmospheres computed under different physical assumptions. The main difference between the types of models is the treatment of geometry and convective energy transport, as described in the subsequent sections.

For the purposes of comparison with previous calculations in the literature \citep{Asplund2004,Amarsi2018}, we employ the solar MARCS model also used in \citet{Bergemann2019}. This is a 1D LTE line-blanketed model atmosphere computed under the assumption of hydrostatic equilibrium. Convective energy transport is parametrised using the mixing-length-theory (MLT), with the MLT $\alpha$ constant set to $0.5$. The micro-turbulence parameter is set to $1$ km\,s$^{-1}$. 

The 3D model atmospheres employed in this paper were computed using two 3D RHD codes. The \texttt{STAGGER} code \citep{Nordlund1995,Nordlund2009} provides a well established 3D radiation-hydrodynamic simulation of sub-surface stellar convection. The solar simulation used in this work \citep{Collet2011, Magic2013} encompasses the entire  photosphere as well as the upper part of the convective zone. \texttt{Bifrost} is a 3D (M)RHD code capable of simulating the magnetic solar chromosphere \citep{2011A&A...531A.154G}. We used a \texttt{Bifrost} simulation 
ranging from $\approx -2.5$ Mm below the optical surface up to $8$ Mm above. From one snapshot of this full-fledged "chromospheric" simulation, we extracted a purely "photospheric" simulation by cutting off the layers above $1$ Mm height. This simulation was given 15 minutes to re-adjust to the new boundary. The horizontal extent of the \texttt{STAGGER} atmosphere is $8$x$8$ Mm, whereas it is $12$x$12$ Mm for the \texttt{Bifrost} atmosphere. Yet both are large enough to host at least $10$ granules at the surface, and have the (x,y,z)-resolutions of 240x240x230 (\texttt{STAGGER}) and 512x512x512 (\texttt{Bifrost}), respectively.
The horizontal resolution of all models was scaled down to allow for repeated NLTE calculations with modest computational cost. Our detailed tests (Fig. \ref{fig:restest}, also see Sect. \ref{sec:errors}) allow us to conclude that (x,y) resolutions in excess of (30,30) are not necessary, as already this configuration preserves the main physical properties of line formation and yields line profiles that differ by no more than $2 \%$ from the full geometric setup. This uncertainty is sub-dominant to other sources of error in the abundance analysis and can be tolerated. Nonetheless, we account for this error in the determination of the solar O abundance (Sect. \ref{sec:likelihood}). The vertical resolution remains unaffected, namely 230 grid-points for the \texttt{STAGGER} simulation and 512 or 192 grid-points for the chromospheric and photospheric \texttt{Bifrost} simulations, respectively. 

\texttt{STAGGER} and \texttt{Bifrost} codes differ in the input abundances and in the equation of states (EOS). \texttt{STAGGER} is using a modified version of the EOS from Mihalas et al. (1988) and abundances from \citet{Asplund2009}. The \texttt{Bifrost} simulation used here was originally made to make comparisons with older simulations from Stein \& Nordlund and used an EOS from \citet{Gustafsson1973}, abundances and continuum opacities from \citet{Gustafsson1975} computed using the Uppsala background opacity package \citep{Gustafsson1973}. 

The temperature structure of the combined down-scaled snapshots of all models is shown in Fig. \ref{fig:STAPH}. The dashed lines present the structure of the spatially and temporally averaged (\tda) model atmospheres. The averaged models were computed by interpolating the full 3D cubes to an equally spaced optical depth scale and averaging over horizontal slices of equal optical depth. Subsequently all variables were averaged temporally over all spatially averaged snapshots. This was done directly for the temperature and velocity field, but in the case of electron density temporal and horizontal averaging was performed on logarithm of electron number density.
\subsection{NLTE statistical equilibrium calculations in 1D and 3D}\label{sec:stateq}

\begin{figure}
    \centering
    \includegraphics[scale=0.7]{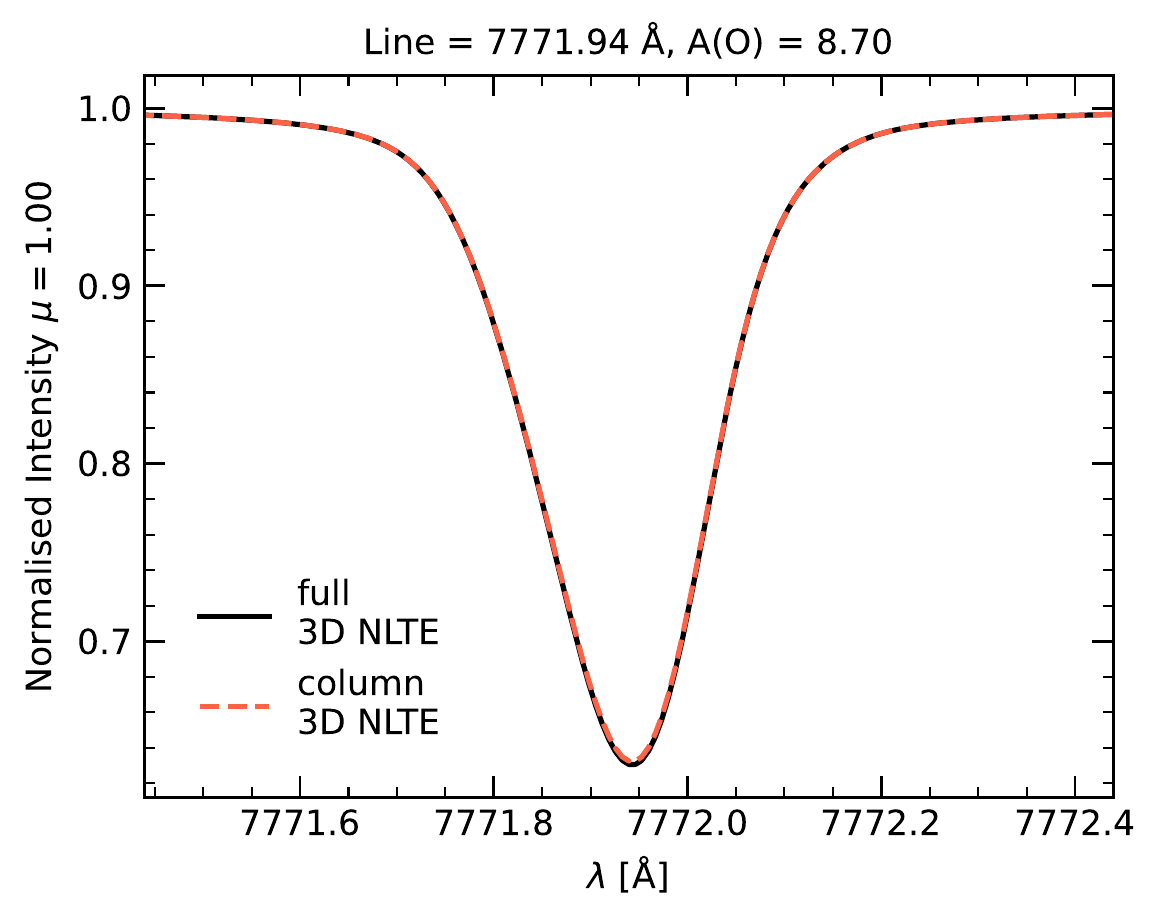}
    \caption{Line profiles of the O I 7771.94 Å line calculated using the column-by-column 3D NLTE solver and the full 3D NLTE radiative transfer solver. The difference in the line EW is only $0.6 \%$.}
    \label{fig:col3D}
\end{figure}

We use two SE codes, MULTI2.3 \citep{Carlsson1992} and MULTI3D \citep{Leenaarts2009, Leenaarts2012}, both updated as described in \citet{Bergemann2019} and \citet{Gallagher2020}. MULTI2.3 solves the detailed equations of SE in a 1D plane-parallel geometry using the Accelerated Lambda Iteration (ALI) method \citep{Rybicki1991, Rybicki1992}. Radiation transfer is solved by the method of long characteristics. All calculations in this work were carried out using the local operator acting on the source function. MULTI3D solves the radiative transfer equations in 3D geometry on a Cartesian mesh via the short characteristics method \citep{Kunacz1988}. It is also capable of solving radiative transfer column-by-column  mode, that is ignoring the influence of slanted rays. Extensive tests of the full 3D and the column-by-column solution revealed that the differences between the both approaches are negligibly small, especially for the Sun, as the line EWs change by less than $0.5 \%$ \citep[see, e.g. the tests carried out by][]{Bergemann2019}. Our calculation for the O lines also shows that the difference is only $0.6 \%$ (Fig. \ref{fig:col3D}). Therefore, we employ the column-by-column radiative transfer solution with MULTI3D in this work. 

In addition to continuum opacities, both MULTI2.3 and MULTI3D are capable of including  background line opacity in the form of a linelist. For O, this not important, because the statistical equilibrium of the element is not sensitive to line blanketing. For \nii, however, a comprehensive treatment of line blanketing across the entire UV and optical wavelength ranges is critical, because the element, similar to other Fe-group species, is photoionisation-dominated. Since MULTI2.3 is also capable of including sampled opacities, we use the code to compute 3D NLTE corrections for Ni separately using the column-by-column solver and apply them to 3D LTE calculations for Ni, in order to get 3D NLTE line profiles of \nii~lines. This approach is needed, because it is currently not feasible to run MULTI3D with a linelist that is comprehensive and detailed enough, that is, that includes the bound-bound transitions of all relevant absorbers across the entire range of frequencies in the Ni model atom. We have verified \citep[see also][]{Gallagher2020}, that both codes provide identical results when the same initial conditions are used. A detailed analysis of the influence of line blanketing in 3D NLTE calculations for different chemical elements will be presented elsewhere (Semenova et al. in prep).
\subsection{Abundance calculations for O and Ni}\label{sec:abundcal}
The calculations of O abundance were carried out using the model atmospheres described in Sect. \ref{sec:atmos}. In the 3D analysis, we employ $13$ consecutive \texttt{STAGGER} snapshots, as well as $10$ consecutive snapshots for each of the two \texttt{Bifrost} model atmospheres (with and without chromosphere). All model atmospheres were probed by $17$ rays at $5$ different angles, one ray being the disc center intensity ($\cos{\theta}=\mu=1$) and the other $16$ rays going north, east, south and west at inclinations $\mu= 0.8, 0.6, 0.4,$ and $0.2$ in order to match the IAG and SST observations.

We use this setup to compute \oi~line profiles for a dense grid of O abundances, sampling the $\logo$ range from $8.50$ to $8.90$ dex in 3D and $8.50$ to $9.20$ dex in 1D.  Since the diagnostic lines of O are not strong, it is sufficient to assume an equidistant abundance spacing of $0.2$ dex. For Ni, we assume the meteoritic abundance of $\logni = 6.23$ dex from \citet{Lodders2003}. This quantity has an uncertainty of $0.04$ dex, and it appears to be superior to the values inferred by spectroscopic methods. \citet{Asplund2009} find the Ni abundance of $6.22 \pm 0.04$ dex, using 3D LTE modelling. This value is significantly higher than the 3D LTE value recommended by \citet{Caffau2015} ($\logni = 6.11 \pm 0.04$ dex). The 1D LTE estimate by \citet{Wood2014}, based on the analysis of 76 \nii~ and \niii~ lines, is $6.27 \pm 0.06$ dex. We adopt the meteoritic Ni abundance in our work, because of the afore-mentioned differences in the photospheric results by different groups. We return to this issue in Sect. \ref{sec:disc}.

The region around the forbidden [O~I] line is modelled by co-adding the profiles of the O and Ni lines in 1D LTE, 1D NLTE, or 3D NLTE respectively and fitting the combined line profiles to the observations. Comparing this approximate approach to the exact procedure, in which the absorption coefficients are co-added instead of the intensity or flux profiles, we find that it provides an almost identical solution (differing in EW by less than $0.5 \%$). Our detailed approach to the model-data comparison and abundance diagnostics is described in Sect. \ref{sec:finalCLV} and \ref{sec:likelihood}. 
\section{Results}\label{sec:results}

In this section, we describe the results of our calculations using different model atmospheres and model atoms. We start with a brief overview of the formation properties of \oi~lines in Sect. \ref{sec:linforO}, continue with the statistical equilibrium of Ni in Sect. \ref{sec:linforNi}, discuss our results for the spatially-resolved spectra in Sect. \ref{sec:finalCLV} and for different atomic models in Sect. \ref{sec:finalatom}, and comment on the relevance of 3D (M)HD in the abundance calculations in Sect. \ref{sec:finalchrom}. We then summarise the methods used in the probabilistic abundance analysis and present the final O abundances in Sect. \ref{sec:likelihood}. Finally, we compare our results with recent estimates in the literature in Sect. \ref{sec:disc} and draw conclusions.

\begin{figure}
\includegraphics[width=0.95\columnwidth]{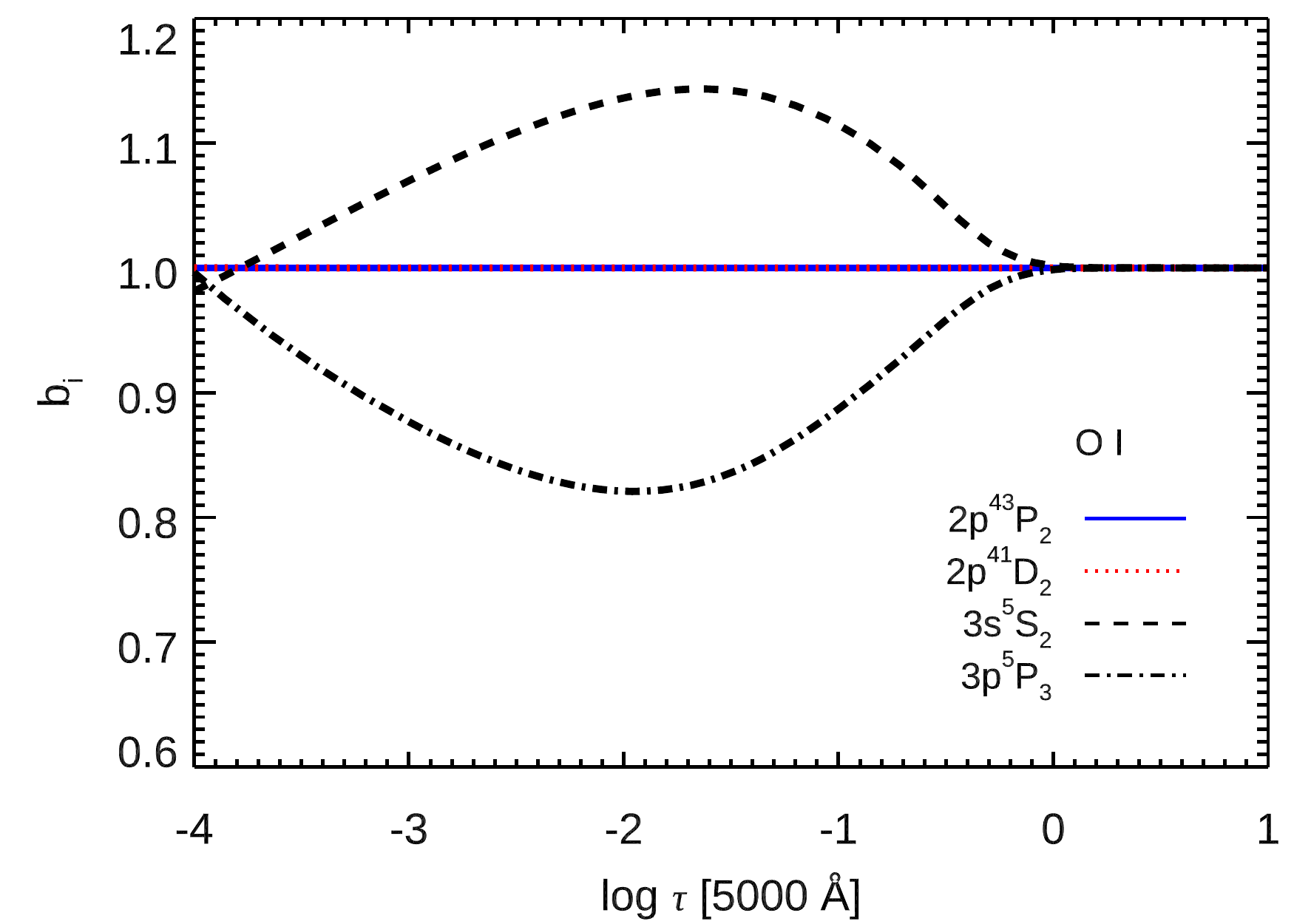}
\caption{Departure coefficients of the energy levels involved in the diagnostic transitions of \oi.}
\label{fig:dep1DO}
\end{figure}
\begin{figure}
\includegraphics[scale=0.65]{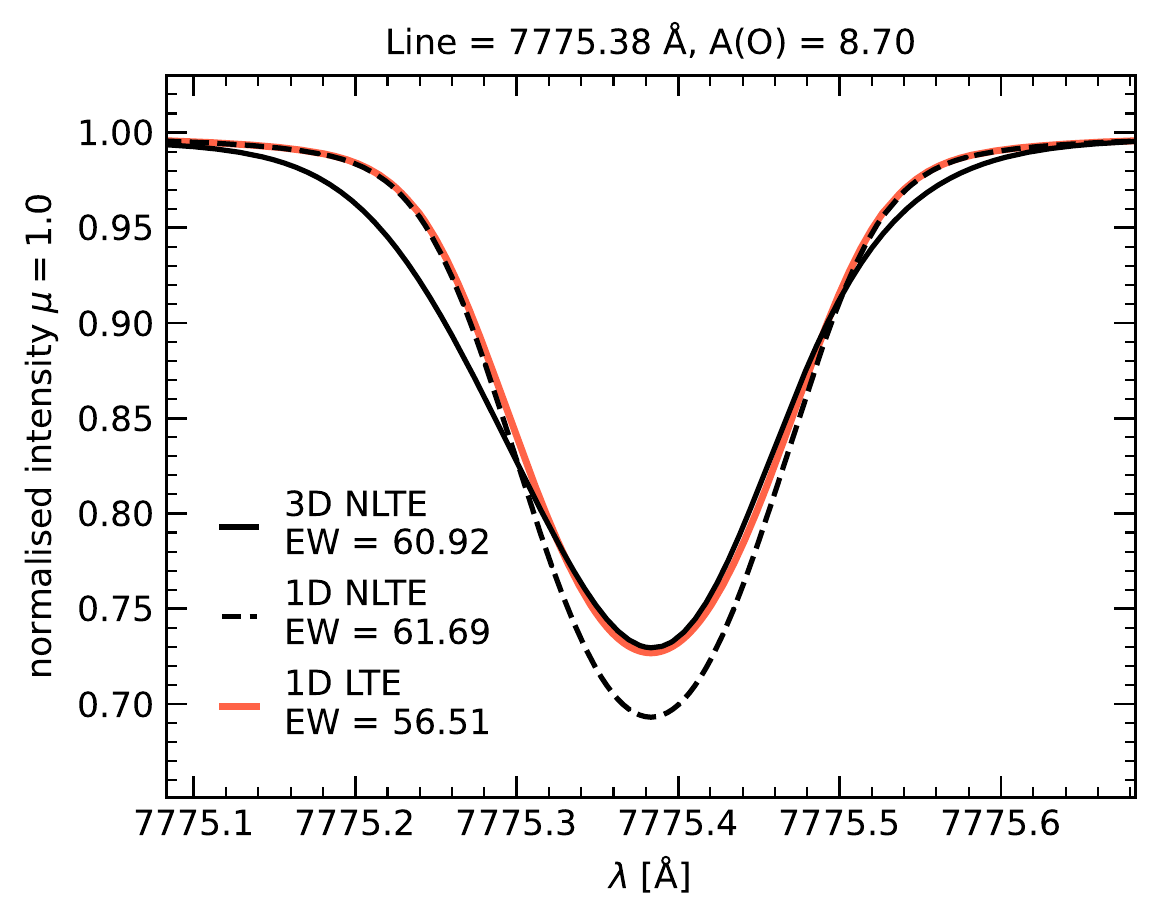}
\includegraphics[scale=0.65]{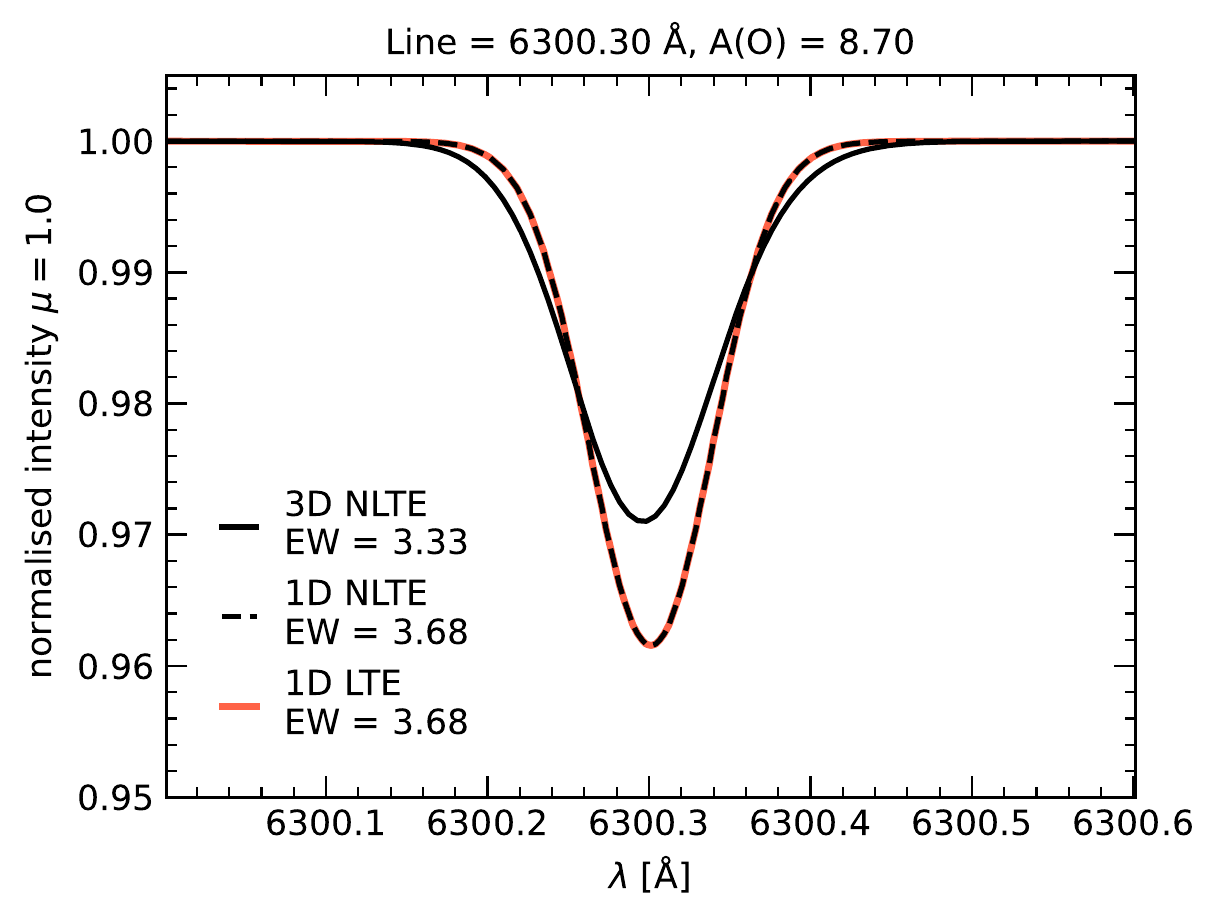}
\caption{Model 1D LTE, 1D NLTE, and 3D NLTE line profiles of the \oi~ lines at 7775.39 \Angstrem (top panel) and 6300.3 \Angstrem (bottom panel).}
\label{fig:lin1DO}
\end{figure}
\subsection{O line formation}\label{sec:linforO}

The formation of O triplet lines at 7771, 7774, and 7775 \Angstrem~has been a subject of intense discussions over the past decades. \citet{Kiselman1991} was among the first to point out that for the solar model atmosphere the NLTE effects in the diagnostic O lines, which connect the two high-excitation states, are relatively insensitive to the detailed structure of the model atom and are controlled by scattering in the lines. The formation of O triplet lines in the solar photosphere was extensively discussed in \citet{Asplund2004}, \citet{Steffen2015} and \citet{Amarsi2018}.

Our calculations confirm that the NLTE effects in the triplet lines are mostly driven by photon losses in the lines \citep{Bergemann2014} that leads to the deviation of the line source function from the Planck function. The overpopulation of the lower level comes at the expense of the ground state population of \oi, via a sequence of transitions that involve charge exchange reactions with \oii, re-combinations to the high-excitation \oi~levels, and spontaneous transitions to the lower levels. This is illustrated in Fig. \ref{fig:dep1DO}, which shows the departure coefficients $b_i$ of the lower \Ot{3s}{5}{S}{o}{2} and upper \Ot{3p}{5}{P}{}{} energy states, which connect the diagnostic triplet lines, against the continuum optical depth at $5000$ \Angstrem. Owing to the small Boltzmann factor, the line formation is restricted to a very narrow range of optical depths ($-1$ < $\log(\tau_{500})$ < 0). In this region, the ratio of the departure coefficients of the upper and lower energy states drops below unity, $b_j/b_i < 1$. As a consequence, the NLTE profiles of the permitted \oi~ lines at 777 nm come out stronger compared to the LTE line profiles (Fig. \ref{fig:lin1DO} top panel). In contrast, the departure coefficients of the levels involved in the transition (the [O~I] line at 630 nm) are very close to unity and the NLTE effects in the line profiles are negligibly small (Fig. \ref{fig:lin1DO}, bottom panel). 
\begin{figure}
\includegraphics[scale=0.7]{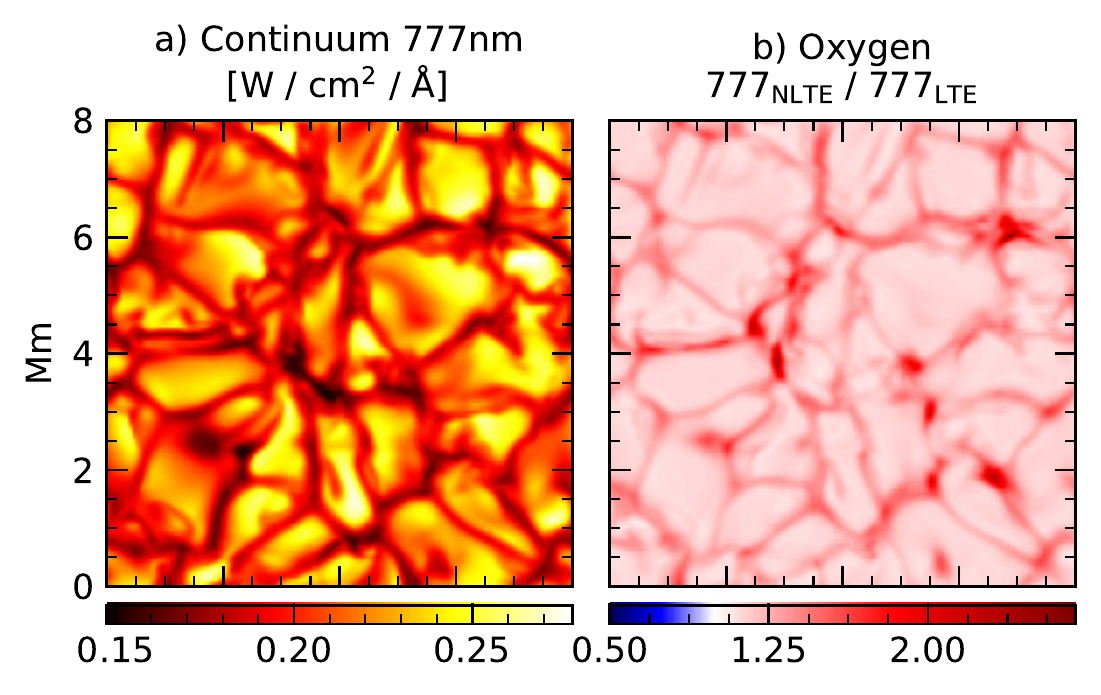}
\includegraphics[scale=0.7]{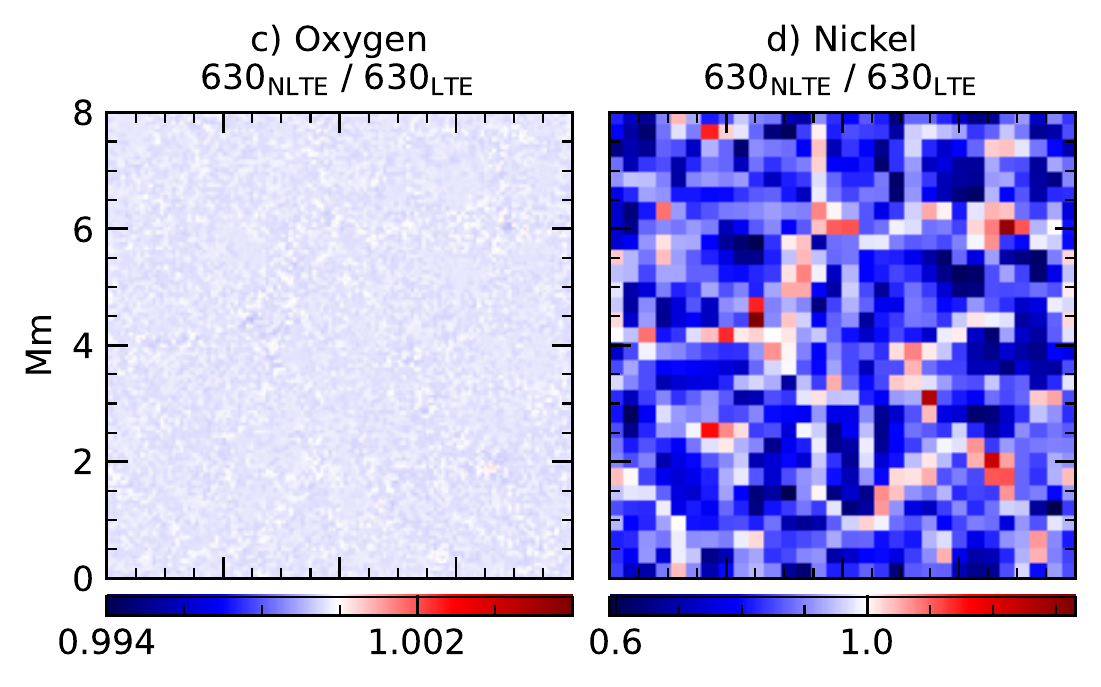}
\caption{Model continuum intensity in the region of 777 nm is shown in panel (a) in absolute units of W/cm$^2$/\Angstrem. Ratios of NLTE to LTE EWs of the oxygen 777 nm triplet lines (panel b), oxygen 630 nm line (panel c) and nickel 630nm line (panel d) computed using the Stagger model atmosphere.}
\label{fig:EWmaps}
\end{figure}

This formation of permitted \oi~ lines is very similar in 1D NLTE and in 3D NLTE \citep[see also][]{Asplund2004,Steffen2015, Amarsi2018}. The 777 nm lines are stronger in 3D NLTE compared to 1D LTE or 3D LTE, that is, the abundances inferred from 3D NLTE line profiles are significantly lower. The forbidden line is barely sensitive to NLTE effects, even in 3D convective models. This is demonstrated in Fig. \ref{fig:EWmaps}, which shows the absolute continuum intensities (panel a) and the ratios of the NLTE and LTE equivalent widths computed using the \texttt{Stagger} 3D model atmosphere for the 777 nm triplet (panel b) and for the [O~I] line at 6300 \Angstrem~ (panel c). The 777 nm lines are stronger in 3D NLTE compared to 3D LTE across the entire simulation surface and the NLTE effects are particularly prominent for the lines that form above the inter-granular lanes. The forbidden [O~I] line forms nearly in LTE, although very weak departures from LTE (line weakening in NLTE) are present above the inter-granular lanes.
\subsection{O NLTE abundance corrections}\label{sec:nltecor}

Table \ref{tab:NLTEcorO} compares the results in terms of 1D NLTE and 3D NLTE abundance corrections, that is the amount by which the abundance of O has be changed in order to match the EWs of 1D NLTE and 3D NLTE line profiles to those of 1D LTE line profiles. These quantities were computed for the disc-center intensity and fluxes using the O abundance of 8.70 dex\footnote{We note that the correction is almost identical, if the abundance of $8.90$ dex is adopted.}. The abundance corrections are not used in the abundance analysis, but are useful for their didactic value. 
%
% O NLTE corrections mu=1
%
%
\begin{table*}
\renewcommand{\footnoterule}{}
\caption{1D NLTE and 3D NLTE abundance corrections for the disc center ($\mu = 1$) intensities and fluxes of the diagnostic lines of O. The model atoms are defined as follows (Sect. \ref{sec:atomO}): (a) LCAO data, (b) QPC data, and (c) LCAO data supplemented with the Kaulakys data.}
\label{tab:NLTEcorO}
\begin{center}
\begin{tabular}{c ccc | ccc | ccc | ccc }
\toprule
 Line & \multicolumn{9}{c}{Intensity ($\mu = 1$)}  & \multicolumn{3}{c}{Flux} \\
      & \multicolumn{3}{c|}{1D NLTE} &  \multicolumn{3}{c|}{<3D> NLTE} &   \multicolumn{3}{c|}{3D NLTE} & \multicolumn{3}{c}{LCAO} \\
 \Angstrem & a &  \textbf{b} &  c &  a &  \textbf{b} &  c &  a &  \textbf{b} &  c & 1D NLTE & <3D> NLTE & 3D NLTE \\
\midrule
   7771.940  & $-$0.20 & $-$0.17  & $-$0.11  & $-$0.16 &  $-$0.13 & $-$0.06 & $-$0.17 & $-$0.15 & $-$0.08   & $-$0.28 & $-$0.27 & $-$0.31 \\
   7774.170  & $-$0.18 & $-$0.16  & $-$0.10  & $-$0.14 &  $-$0.11 & $-$0.05 & $-$0.15 & $-$0.13 & $-$0.07   & $-$0.26 & $-$0.24 & $-$0.29 \\
   7775.390  & $-$0.16 & $-$0.14  & $-$0.08  & $-$0.11 &  $-$0.08 & $-$0.02 & $-$0.14 & $-$0.12 & $-$0.07   & $-$0.23 & $-$0.20 & $-$0.27 \\
   6300.304  & ~~0.00  &  ~~0.00  &  ~~0.00  & $+$0.04 &  $+$0.04 & $+$0.04 & $+$0.05 & $+$0.05   & +0.05   &  ~~0.00  & $+$0.04 & $+$0.04 \\
\bottomrule
\end{tabular} 
\end{center}  
\end{table*}   

Not surprisingly, the NLTE results and, consequently, the NLTE abundance corrections depend strongly on the input atomic data, foremost on the rates of H impact transitions. Using the quantum-mechanical data (LCAO or QPC), we obtain the 1D NLTE corrections in the range from $-0.20$ dex (7771 \Angstrem~line) to $-0.16$ dex (7775 \Angstrem~line) for the LCAO model, and $-0.17$ dex (7771 \Angstrem~line) to $-0.14$ dex (7775 \Angstrem~line) for the QPC model. However, the results change dramatically, if we include the Kaulakys data in the model, by co-adding them with the LCAO datasets. In this case, the 1D NLTE corrections do not exceed $-0.11$ dex for the 7771 line and $-0.08$ dex for the 7775 \AA~line. All other quantities in the atomic models, such as the representation of H-impact transition rates for the high-excitation energy states, electron collisions, or photo-ionisation, do not influence the results at any significant level. Likewise, the NLTE level populations and the profiles of the diagnostic lines remain nearly identical, regardless of whether Drawin's rates or a blanket constant rate coefficient of $10^{-20}$ are assumed for the majority of the uppermost states in \oi~(Fig. \ref{fig:DrScal} in Appendix). As emphasised, however, it is important to include these values to ensure rapid convergence, which is important especially in extremely time-consuming 3D NLTE calculations.

The NLTE corrections computed using the spatially- and temporarily-averaged $<$3D$>$ Stagger model atmosphere are less extreme. For the LCAO and QPC atomic models (no Kaulakys data), the largest $<$3D$>$ NLTE corrections do not exceed $-0.16$ dex and $-0.13$ dex (the 7771 \AA~line), respectively. The model atom that includes LCAO and Kaulakys data returns the results that are not too different from LTE: the largest NLTE correction amounts to $-0.06$ dex (7771 \AA~line), whereas the 7775 \AA~line forms nearly in LTE, with the NLTE correction of only $-0.02$ dex.

Interestingly, the 3D NLTE abundance corrections are very close to the $<$3D$>$ NLTE results, although the former are slightly larger in absolute value. For the LCAO model atom, the 3D NLTE corrections amount to $-0.17$ dex for the 7771 \Angstrem~line, $-0.15$ dex for the 7774 \Angstrem~line, and $-0.14$ dex for the 7775 \Angstrem~line. The latter line, which is the weakest of all three, is most sensitive to the structure of the model atmospheres and to the properties of the model atom. Its NLTE correction changes from $-0.16$ to $-0.02$, depending on the details of modelling. The forbidden oxygen line forms almost in LTE, but it shows a non-negligible sensitivity to convection. The 3D NLTE correction for the 630 nm [O~I] line amounts to $+0.05$ dex relative to 1D LTE.

Table \ref{tab:NLTEcorO} also shows our 1D NLTE and 3D NLTE corrections for fluxes computed using the LCAO model atom. These results can be directly compared to the corresponding values from \citet[][their Tab. 2]{Asplund2004}. In 1D NLTE, they obtain $-0.24$ dex for the 7771 \Angstrem~line, $-0.23$ dex for the 7774 \Angstrem~line, and $-0.20$ dex for the 7775 \Angstrem~line. In 3D NLTE, they derive $-0.27$ dex for the 7771 \Angstrem~line, $-0.24$ dex for the 7774 \Angstrem~line, and $-0.20$ dex for the 7775 \Angstrem~line. Our values are in a good agreement with this study, although the amplitude of our NLTE corrections is slightly larger. This can be explained by the complexity of the atomic model. Our model includes 120 states of \oi~coupled by a variety of processes (Sect. \ref{sec:atomO}, whereas the atomic model employed by \citet{Asplund2004} includes only $20$ \oi~states coupled by 43 radiative b-b transitions and electron collisions, and it neglects H-impact collisions. Since the model atom from that study was not published, to check the influence of model atom completeness we performed a sequence of calculations in 1D NLTE (using the MARCS and $<$3D$>$ model atmosphere), progressively reducing the complexity of the model atom down to $22$ levels coupled by $57$ radiative b-b transitions. We find that the model that includes $41$ oxygen states, with \oi~closed by the 4f $^3$F term (102968 cm$^{-1}$) reproduces exactly the results obtained using the complete model. However, atomic models \textit{smaller} than this limit lead to \textit{more modest} NLTE corrections. A 22-level atom underestimates the NLTE corrections by $0.07$ dex, resulting into 1D NLTE corrections of the order $-0.21$ (7771) to $-0.16$ dex. The residual differences can be explained by \citet{Asplund2004} not including H-impact collisions in their model and using the older version of the Stagger model (M. Asplund priv. comm).

In summary, our findings are similar to those of the previous studies of 1D NLTE and 3D NLTE line formation of \oi~lines in the solar atmosphere. As we will show below, however, the details of the \oi~model atom (primarily, the H collision data) and the approach to the statistical analysis of spectral lines, lead to quantitatively different results, when it comes to the analysis of the solar photospheric O abundance.
\subsection{Ni line formation}\label{sec:linforNi}
The statistical equilibrium of Ni was previously investigated by \citet{Bruls1993}. Using a model atom with $19$ energy levels and $81$ radiative transitions, they found that \nii is subject to over-ionisation and line pumping, the processes that generally favour under-population of energy levels and cause weakening of spectral lines compared to LTE. Over-ionisation and pumping are driven by strong non-local UV radiation field and, similar to other Fe-group species \citep{Bergemann2012}, influence the levels with excitation energies of $\sim 3$ to $4$ eV. \citet{Ding2002} used a simplified 10-level version of the model developed by \citet{Bruls1993} in order to study the NLTE line formation properties of the at $6767.8$ \Angstrem~ line, which is typically used in helioseismology. \citet{Vieytes2013} developed a larger model atom, including $61$ energy levels and $401$ radiative transitions, and applied the model atom in the calculations of the solar broad-band spectrum. None of these studies considered the formation of the \nii~feature at 6300.341 \Angstrem~that contaminates the [O~I] line.

\begin{figure}
\includegraphics[width=0.95\columnwidth]{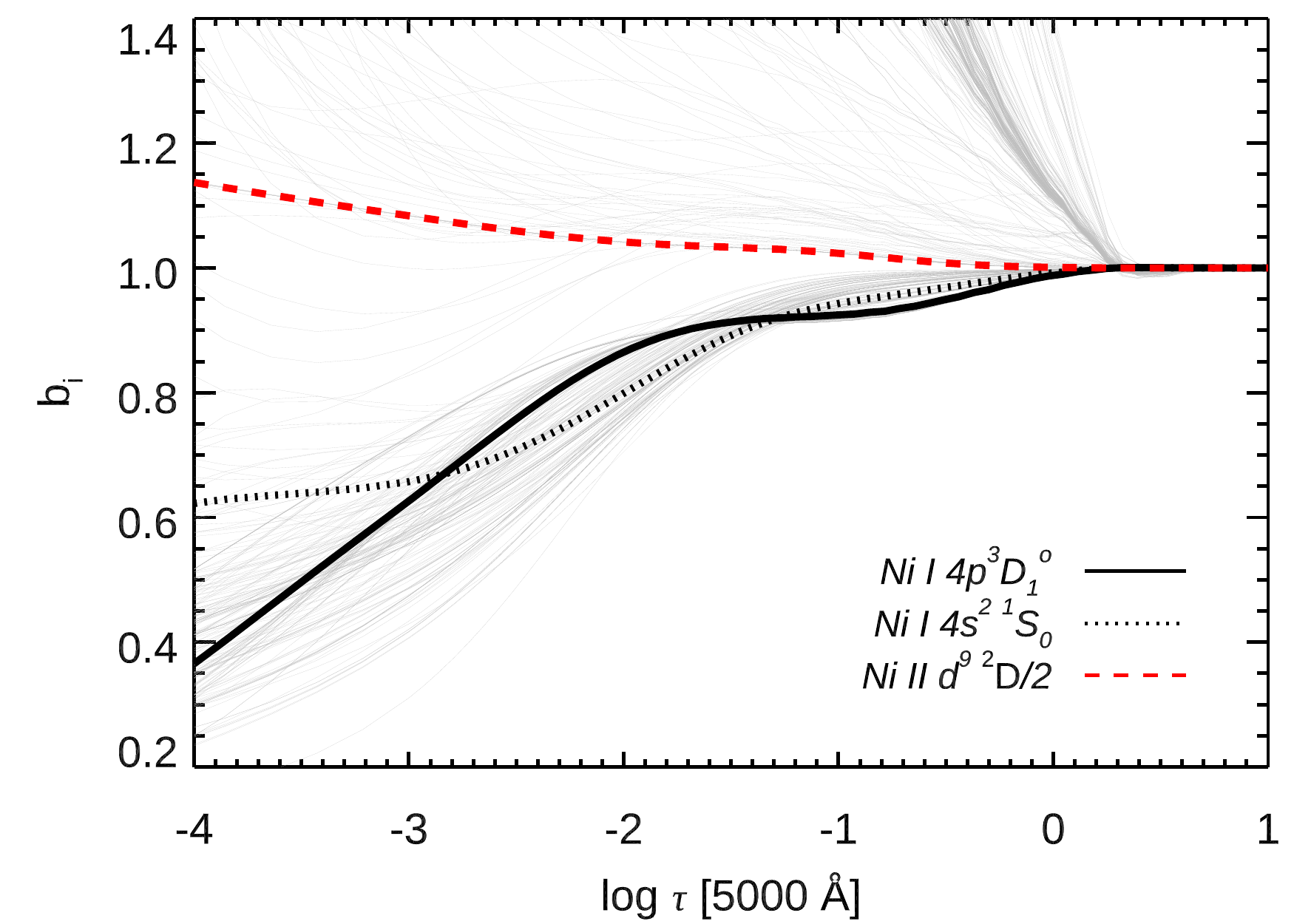}
\caption{NLTE departure coefficients of energy levels in \nii. The levels involved in the 6300.341 \Angstrem~transition are highlighted.}
\label{fig:dep1DNi}
\end{figure}

Our calculations confirm that \nii~behaves as a classical minority element with numerous important ionisation edges in the UV, which is subject to over-ionisation by strong non-local radiation field. As seen in the diagram of the departure coefficients (Fig. \ref{fig:dep1DNi}), the majority of energy levels in \nii~ are under-populated. The states with excitation energies of $2$ to $4$ eV are over-ionised, whereas the higher-lying \nii~states are collisionally coupled with the latter and therefore display similar NLTE effects. The ground state of \niii~ is nearly thermalised across the entire optical depth scale, but higher-lying \niii~states are over-populated, closely resembling the behaviour seen in other Fe-group elements, such as Cr, Ti, Mn, and Co. 
\begin{table}
\renewcommand{\footnoterule}{} 
\caption{1D and 3D NLTE abundance corrections (in dex) for the disc center intensities of the Ni I line at 6300.341 \AA. S$_{\rm bf}$ the scaling factor to the collisional ionisation rates computed using the Drawin formula.}
\label{tab:NLTEcorNi}
\begin{center}
\begin{tabular}{c c ccc ccc}
\toprule
\multirow{2}{*}{Line} &  \multicolumn{1}{c}{LTE} & \multicolumn{6}{c}{NLTE} \\ 
\cmidrule(lr){3-8}
&  3D    & \multicolumn{3}{c}{1D} & \multicolumn{3}{c}{3D} \\
\cmidrule(l){3-5} \cmidrule(l){6-8}
  &  & $0$ & $100$ & $1000$ & $0$ & $100$ & $1000$ \\
\midrule
6300.341 &  0.06  &  0.15  &  0.08 &  0.02 & 0.21 & 0.14 & 0.08 \\
\bottomrule
%\hline
%\hline        
\end{tabular} 
\end{center}  
\end{table}

\begin{figure}
\includegraphics[scale=0.7]{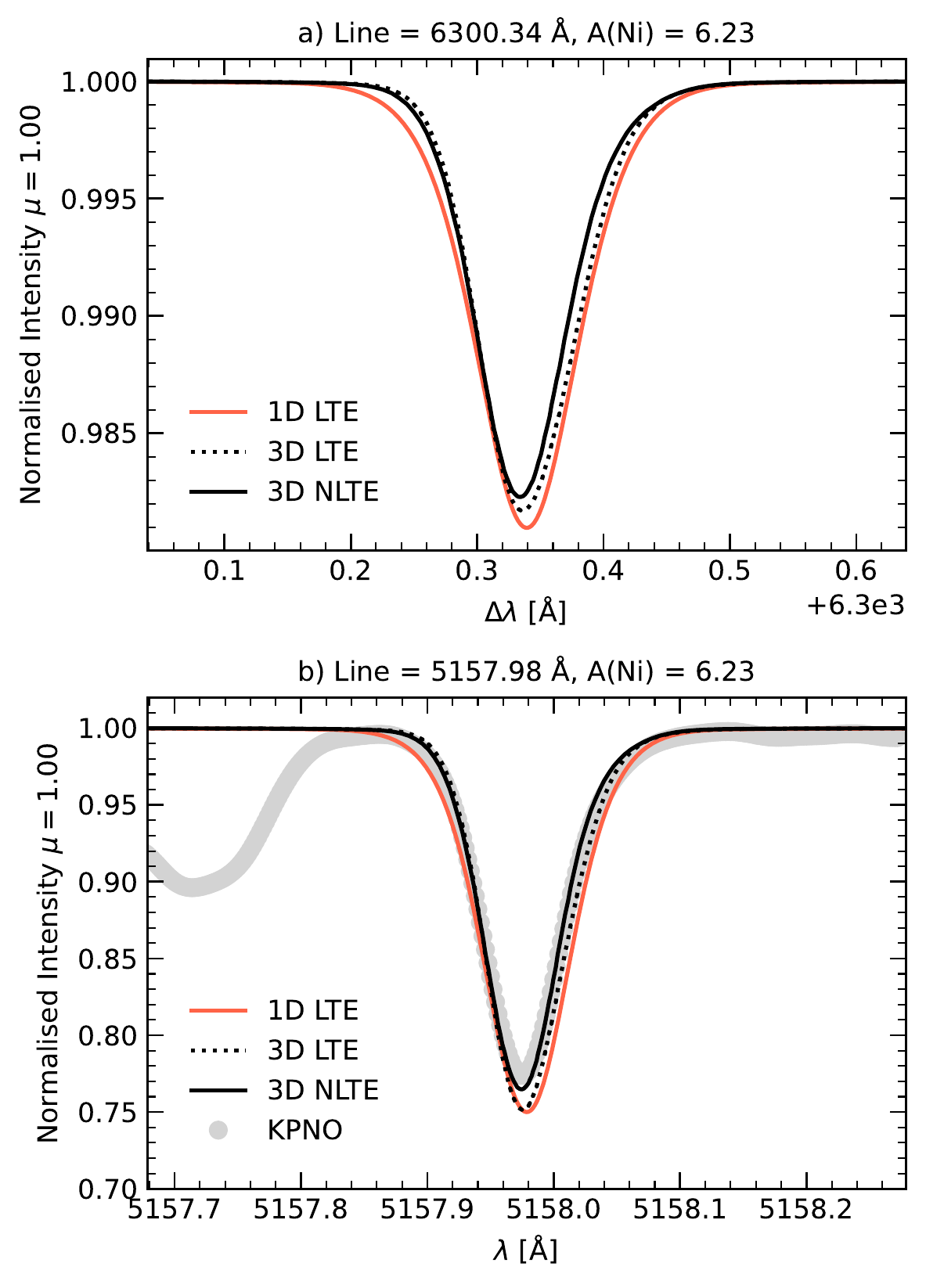}
\caption{1D LTE, 1D NLTE, and 3D NLTE line profiles of the Ni lines at 6300.34 \Angstrem~(top panel) and 5157.98 \Angstrem~(bottom panel). The 6300.34 \Angstrem~feature blends with the  forbidden O I line at 6300.304 \Angstrem.}
\label{fig:lin1DNi}
\end{figure}

Table \ref{tab:NLTEcorNi} presents the 3D LTE, 1D NLTE, and 3D NLTE corrections for the key Ni line at 6300.34 \AA. The line is clearly sensitive to convection, with the 3D LTE estimate being higher by $0.06$ dex compared to 1D LTE. The NLTE values were computed using three atomic models of Ni, which differ only in the scaling factor to Drawin's ionisation rates. For the excitation rates, we used the scaling factor of 0.05 (Sect. \ref{sec:atomNi}). Clearly, assuming no H-impact ionising collisions (S$_{\rm{bf}} = 0$) or increasing them by a factor of $100$ (S$_{\rm{bf}} = 100$) has a significant influence on the NLTE correction, which changes from $+0.15$ dex to $+0.08$ dex (MARCS) and from $+0.21$ dex to $+0.14$ dex (3D Stagger) respectively. Assuming very large - likely unrealistically too large - Drawin's collisional rates (S$_{\rm{bf}} = 1\,000$), the difference between 3D NLTE and 3D LTE effectively vanishes. The scaling factors of S$_{\rm{bf}} = 0$ and S$_{\rm{bf}} = 1\,000$ provide the maximum (hereafter, Ni-max) and minimum (hereafter, Ni-min) limits on the NLTE effects in Ni lines, respectively. The former approach assumes no H-impact collisional ionisation and the latter results in nearly thermalised (LTE) level populations. It has to be kept in mind, though, that the NLTE effects in Ni are very likely larger and thus, the O abundance might increase. We include the uncertainty related to the amplitude of NLTE effects in Ni into the final estimate of O abundance (Sect. \ref{sec:atomNi}).

Fig.~\ref{fig:lin1DNi} shows the line profiles of the 6300~\AA~ and 5157.98~\AA~\nii~line, computed using the model atom with collisions scaled by S$_{\rm{bf}} = 1000$ and $\logni = 6.23$ dex. The latter spectral line was used in the analysis of the solar Ni abundance in \citet{Scott2015} and it appears to be relatively unblended compared to other \nii~features. Both lines are clearly weaker in 3D LTE compared to 1D LTE, and in 3D NLTE the difference is even larger. The weakening of Ni I lines in 3D LTE was also demonstrated by \citet{Scott2015}, although no NLTE calculations were performed in that study. To the best of our knowledge, \citet{Bruls1993} and \citet{Vieytes2013} are the only studies of NLTE effects \nii~in the solar spectrum, but they do not provide estimates of line EWs and NLTE abundance corrections in 1D and 3D, so a more quantitative comparison cannot be carried out.

Finally, we note that for \nii, no quantum-mechanical estimates of photo-ionisation cross-sections, H and e$-$ collision rates are available. So we have to resort to classical recipes, but we caution that these recipes tend to under-estimate the effects of over-ionisation and line pumping, as detailed NLTE studies of other similar species demonstrated (e.g. Mn: \citealt{Bergemann2019}; Ti: \citealt{Sitnova2020}). Therefore our present calculations can be viewed as a conservative scenario, yet the actual NLTE effects in the \nii~lines are expected to be larger.
\begin{figure*}
\centering
\includegraphics[scale=0.7]{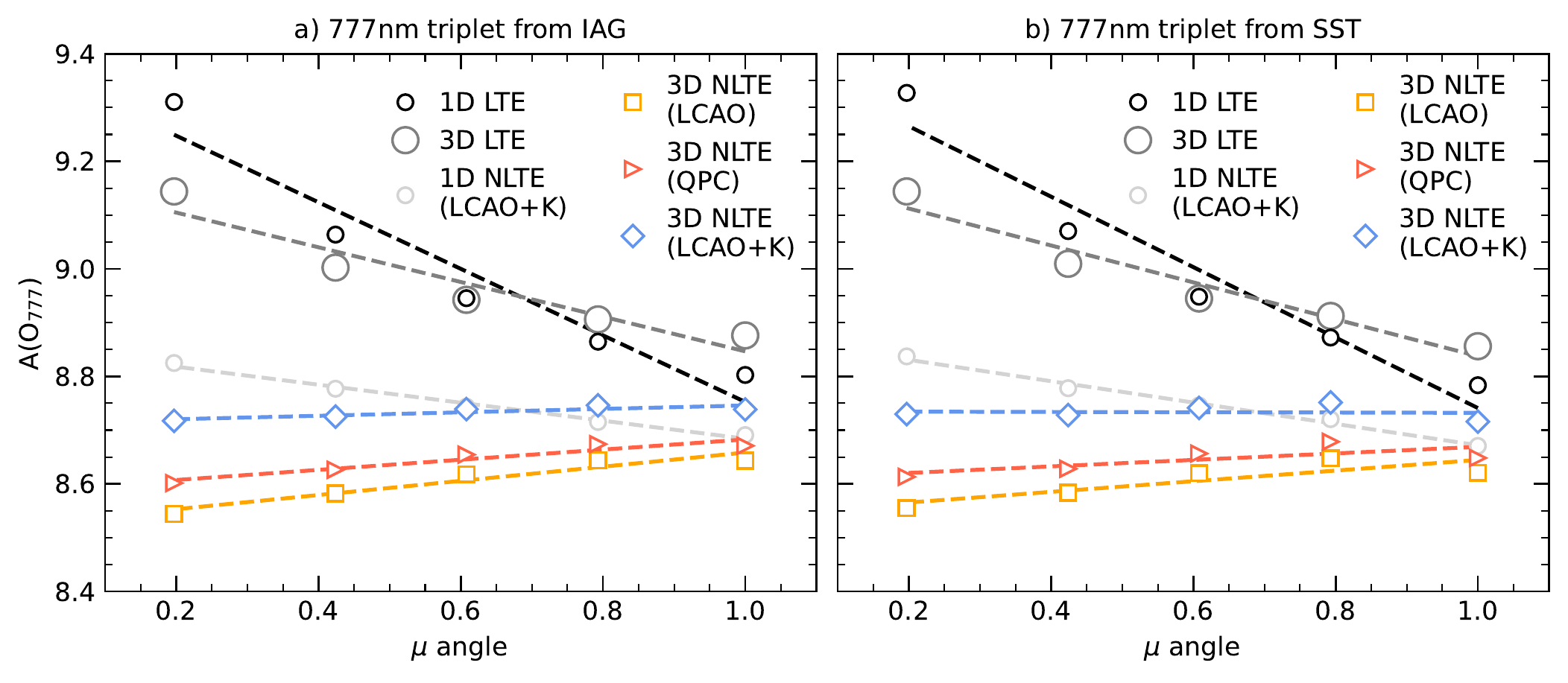}
\includegraphics[scale=0.7]{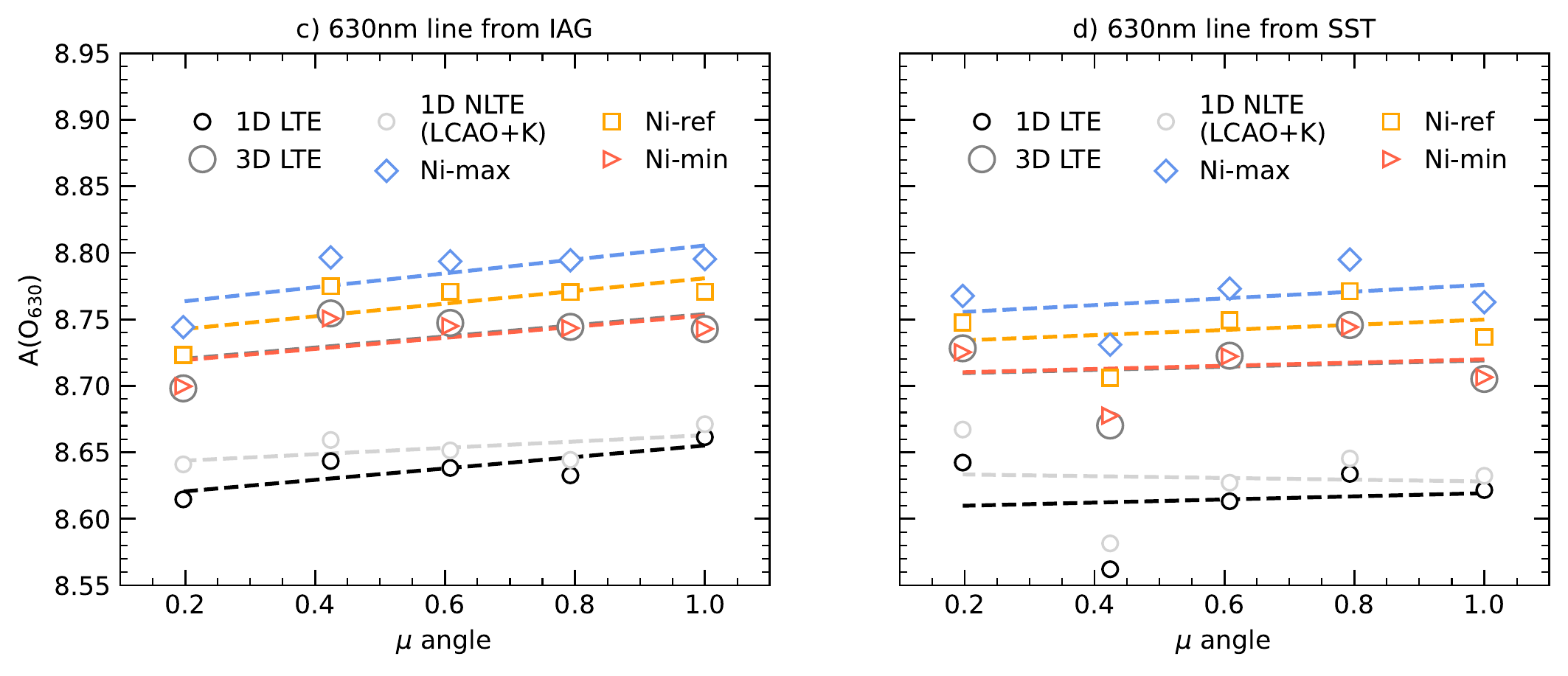}
\caption{Center-to-limb variation of the combined oxygen triplet for different model atom/atmosphere combinations. 1D is represented by the Marcs solar model and 3D is represented by the Stagger solar model.}
\label{fig:CLVall}
\end{figure*}
\subsection{Center-to-limb variation of oxygen abundances}\label{sec:finalCLV}
We begin with the analysis of O abundances obtained for different $\mu$ angles and for different diagnostic lines of O. The best-fit 3D NLTE and 1D LTE models are compared with the spatially-resolved observations of the Sun in Fig. \ref{fig:CLVStaggerLCAO}.

Figure \ref{fig:CLVall} shows our abundance estimates obtained in 1D LTE, 1D NLTE, and 3D LTE for each of the $5$ studied inclinations in the IAG and SST data (Sect. \ref{sec:observations}). Ideally the slope $\alpha$ of the overdrawn trend line would equal zero, meaning that a given model returns self-consistent results at each angle. Low $\alpha$ values are therefore, to a first order, a good indicator of a more reliable result. Additionally, Fig. \ref{fig:CLVall1} (Appendix) shows the observed EWs against the model predictions for 1D LTE, 1D NLTE, and 3D NLTE (LCAO$+$K). This figure is useful for its pedagogical content, to illustrate the failure of standard 1D or 3D LTE modelling. However, we do not use the actual observed EWs in the abundance analysis (see Sect. \ref{sec:likelihood}).

Clearly, the 1D LTE assumption does not yield reliable abundance estimates for the \oi~ triplet (Fig. \ref{fig:CLVall}, top panels), as there is a strong negative correlation between the individual 1D LTE abundance estimates and the viewing angle, ranging from $\logo \approx 8.8$ dex at the disc center to $\logo \approx 9.3$ dex at the limb ($\mu = 0.2$). Also the 3D LTE results are sub-optimal: the slope is slightly smaller compared to 1D LTE, nonetheless the abundances range from $\logo \approx 8.85$ dex at $\mu = 1$ to $\approx 9.10$ dex at $\mu = 0.2$, with a significant systematic bias. 1D NLTE results are similar to 3D LTE in terms of the $\mu$ - $\logo$ slope, however, the disc-center 1D NLTE abundance (here shown for the model atom LCAO$+$Kaul), $\logo = 8.66$ dex, is in a much better agreement with the 3D NLTE disc-center abundance obtained using the \texttt{Stagger} model, $\logo = 8.70$ dex. However, the 3D NLTE approach clearly outperforms the analyses in that it leads to  significantly stronger profiles of the triplet lines at the limb, thereby improving the agreement with the observations \citep[see also][]{Kiselman1995, Asplund2004}. The results obtained using the IAG and SST data are similar in terms of their CLV abundance slopes.

The forbidden [O~I] line (Fig. \ref{fig:CLVall}, bottom panels) is less sensitive to the model atmosphere and to the O model atom, as the angle-dependent abundances are similar in 1D LTE and in 1D NLTE. However, all 3D NLTE and 3D LTE values are about $0.10$ dex higher compared to 1D results. The  differences between various 3D NLTE CLV results are entirely due to the differences in the predicted strength of the Ni blend at 6300.34 \AA. The SST results show peculiar undulations, with abundances at $\mu = 1$ and $\mu = 0.4$ being significantly lower compared to other angles. The origin of this systematic is difficult to pin-point. It is possible that the subtraction of fringing and other instrumental artefacts in these data lead to residual systematic biases. In fact, it is clearly seen in the Fig. 2 of \citet{Pereira2009a} that the O~I line unfortunately coincides with a minimum in the fringe pattern used to model the continuum of the SST data. The O abundances based on the IAG data at different angles are more internally consistent, supporting the choice of these observations as currently the most reliable dataset.
\subsection{Influence of O+H collisions}\label{sec:finalatom}

It is useful to look more closely into the abundances computed from different NLTE model atoms of O. In this respect, the most critical reactions are inelastic collisions between O and H atoms, which have a direct influence on the statistical equilibrium of \oi.

As shown in Sect. \ref{sec:hcol}, our new rate coefficients that describe the transitions in \oi~caused by collisions with H are different from the previous estimates by \citet{Barklem2018}. The latter quantities (LCAO) are on average smaller compared to the data computed using the QPC method. However, if we co-add the data from \citet{Barklem2018} with the data from \citet{Kaulakys1991}, in order to account for the lack of short-rang interactions in the LCAO approach \citep[as recommended by][]{Amarsi2018}, the net result is that for the majority of energy levels the combined LCAO$+$Kaulakys rates are higher compared to the LCAO and to the QPC data. 

As a result, the O abundances computed from the 777 triplet lines using the three atomic models show a clear systematic difference (Fig. \ref{fig:CLVall}, top panels). The LCAO$+$Kaulakys model atom leads to a $\approx +0.1$ (disc center) to $\approx +0.15$ dex (limb) higher solar O abundance compared to the LCAO and QPC models. This is because the latter two models, owing to the less efficient collisional thermalisation they produce, yield larger NLTE effects in the 777 nm \oi~lines. However, neither of the atomic models can be given a preference, based on the CLV - abundance slope, because the moduli of the slopes are very similar. This suggests that the CLV of O lines alone is not sufficient as a metric to distinguish between the three NLTE atomic models. In other words, each of the model atoms - LCAO, QPC, or LCAO$+$Kaulakys - allows us to achieve internally consistent results, which are, however, systematically different with respect to each other.

For completeness, we note that the choice of collisional data in the O model atom  is of no relevance in modelling the 630 nm oxygen line, as the populations of its energy levels are very close to thermal and the NLTE effects are very minor.
\subsection{Model-data comparison}\label{sec:likelihood}
We determine the best-fit abundances by employing the $\chi^2$ statistics, comparing the observed data with the model line profiles. Line profiles for a fine grid of abundances are produced and for each observation we find the most probable abundance by minimising the reduced $\chi^2$:
\begin{equation}
\label{eqn:red_gaussian}
\chi^2 =\frac{1}{N_{\rm pix} - N_{\rm free}}\sum_i^{N_{\rm pix}}  \left[\frac{f_{\rm obs, i}-f_{\rm model, i}}{\sigma_{\rm obs, i}}\right]^2,
\end{equation}
where $N_{\rm pix}$ is the number of wavelength points and $N_{\rm free}$ is 1 as the oxygen abundance is the only free parameter. The observational error $\sigma_{f_{\rm obs}}$ represents the uncertainty caused by the limited signal-to-noise ratio of the data.

For the 777 nm triplet lines, we mask the observations to the region of $\pm$0.5 Å from the line centers and a few surrounding blends (Cr 1 at 7771.76 \Angstrem, Ti 1 at 7771.41 \Angstrem, Ti 1 at 7772.16 \Angstrem, and Sc 1 7772.56 \Angstrem), modelled in 1D LTE, are subtracted from the observations. These blends are weak and they are entirely contained in the absorption features associated with the triplet lines. Since the exact position of the line centers varies between different atlases and $\mu$-angles, the models are slightly blue- or red-shifted, so that the synthetic line centers match the observed data.

The 630 nm [O~I] line is located in a region, which contains a number of blends, Si 1 ($6299.599$ \Angstrem), Sc 2 ($6300.698$ \Angstrem), and Fe 1 ($6301.500$ \Angstrem). However, models are woefully inadequate to reproduce the observations in this region (Fig. \ref{fig:blends}). So to estimate the contribution due to blends, that is to de-blend the O line, we adopted the following approach. We fitted arbitrary Voigt profiles to each of the blending features, allowing for the variation of the amplitude, $\gamma$ and $\sigma$ parameters in the Gaussian and Lorentzian profiles. We also allowed for parabolic bisectors in order to deal with the asymmetry of both lines. This fit (Fig. \ref{fig:deblending}) was then used to determine the O abundance from the de-blended O$+$Ni feature using a mask with the width of $\pm 0.3$ Å. The [O~I] line was modelled using a grid of oxygen abundances, while the Ni was modelled with a fixed abundance of 6.23 dex. The meteoritic value has an uncertainty of $0.04$ dex, but this uncertainty is entirely covered by the difference in the Ni blend contribution caused by adopting different NLTE atomic models of Ni.
\subsection{Systematic errors}\label{sec:errors}

The model (systematic) error $\sigma_{f_{\rm model}}$ comprises several sources, including the errors caused by the oscillator strengths and damping constants, as well as the errors associated with the limited geometric resolution of the 3D model atmosphere. These errors are not included in the $\chi^2$ calculations (Eq. \ref{eqn:red_gaussian}), but are considered separately in the abundance analysis (see Table \ref{tab:errors}).
\subsubsection{Atomic data}\label{sec:fval}

The error caused by the uncertainty of oscillator strengths is fully correlated with the abundance error. As described in Sect. \ref{sec:atomO}, we use the \citet{Hibbert1991} data for the 777 triplet lines and the \citet{Storey2000} data for the 630 nm forbidden line. However, recently new $f$ values for allowed transitions were presented by \citet{Civis2018}. These data are based on the Quantum Defect Theory (QDT) approach and, for the 777 nm triplet lines they are 12\% ($-0.05$ dex) lower than the $f$-values provided by \citet{Hibbert1991} and recommended by NIST. The difference between the $f$-values from the two sources is four times the NIST estimated uncertainty in the transition.

In order to understand the uncertainties of the $f$-values we carried out new atomic calculation. We used the code AUTOSTRUCTURE \citep{Badnell2011} to do multiple calculations of increasing complexity. We started with the same configuration expansion as used by \citet{Storey2000}. Further, we added additional configurations with the same orbitals with principal quantum number n$\le 4$ and configurations with n=5. In each case we tried different orbital optimization schemes. From the different calculations we find a scatter in the $f$-value of the 777~nm transitions of about $10\%$ with a tendency for the rate to be closer to lower value of \citet{Civis2018} than to the \citet{Hibbert1991} value. Therefore, for the 777 lines we adopted the average of both values as our central value, and the associated $f$-value error (2 $\sigma$) is assumed to be the difference between the both quantities.

The $f$-value for the magnetic dipole line at 630~nm is much more difficult to pin point. Such weak lines are generally very sensitive to cancellation effects among different contributions from configuration interaction representations of the atomic levels. The main challenge is not the convergence of the expansion, but the sensitivity to how the orbitals are optimized. The NIST recommended transition probability for this line is $5.63\times 10^{-3}$~s$^{-1}$ with the uncertainty flag B$+$ (better than $7 \%$), which is adopted from \citet{Baluja1988} and \citet{Fischer1983}. The most recent calculations by \citet{Storey2000} yield $6.45\times 10^{-3}$~s$^{-1}$. 
This is a difference of ~$\sim 15\%$, which is more than twice the estimated uncertainty in this rate by NIST. The results of our own calculations show the rate varies widely between $\sim 3\times 10^{-3}$~s$^{-1}$ and $\sim 1\times 10^{-2}$~s$^{-1}$, with the best value being $\sim 8\times 10^{-3}$~s$^{-1}$. As stated earlier, we use the \citet{Storey2000} rate in our analysis, however, to account for the extreme sensitivity of the $f$-values to the details of orbital optimisation, we assume a more realistic uncertainty of 20\%.

Finally, we investigated the uncertainty associated with damping, by performing spectrum synthesis calculations with a modified, by $\pm 10 \%$, value of the damping constant. This, however, has a negligible influence on the O triplet and [O~I] lines and is therefore neglected.
\subsubsection{Geometric resolution}\label{sec:finalgeom}
We also take into account a systematic error associated with the limited geometric resolution of 3D model atmospheres used in radiative transfer calculations. This correction is estimated as follows. The two families of photospheric 3D atmospheres employed in this paper have (x,y,z)-resolutions of 240x240x230 (\texttt{STAGGER}) and 512x512x192 (\texttt{Bifrost}-phot). We restrain from re-sampling the vertical resolution, but down-sample the number of vertical columns, ensuring that we have enough columns to maintain the ratio of intergranular lanes to granules covering the entire simulation. We converted a single \texttt{STAGGER} snapshot to horizontal resolutions 5x5, 10x10, 20x20, 30x30 and 120x120 and computed equivalent widths of the diagnostic \oi~lines for each down-sampled snapshot. Similarly, equivalent widths for a photospheric \texttt{Bifrost} snapshot with different horizontal resolutions between $10$x$10$ and $160$x$160$ were computed. These calculations were carried out in LTE, because our previous analysis \citep{Bergemann2019} confirmed that the scaling is very similar in LTE and in NLTE. 
\begin{figure}
\includegraphics[scale=0.7]{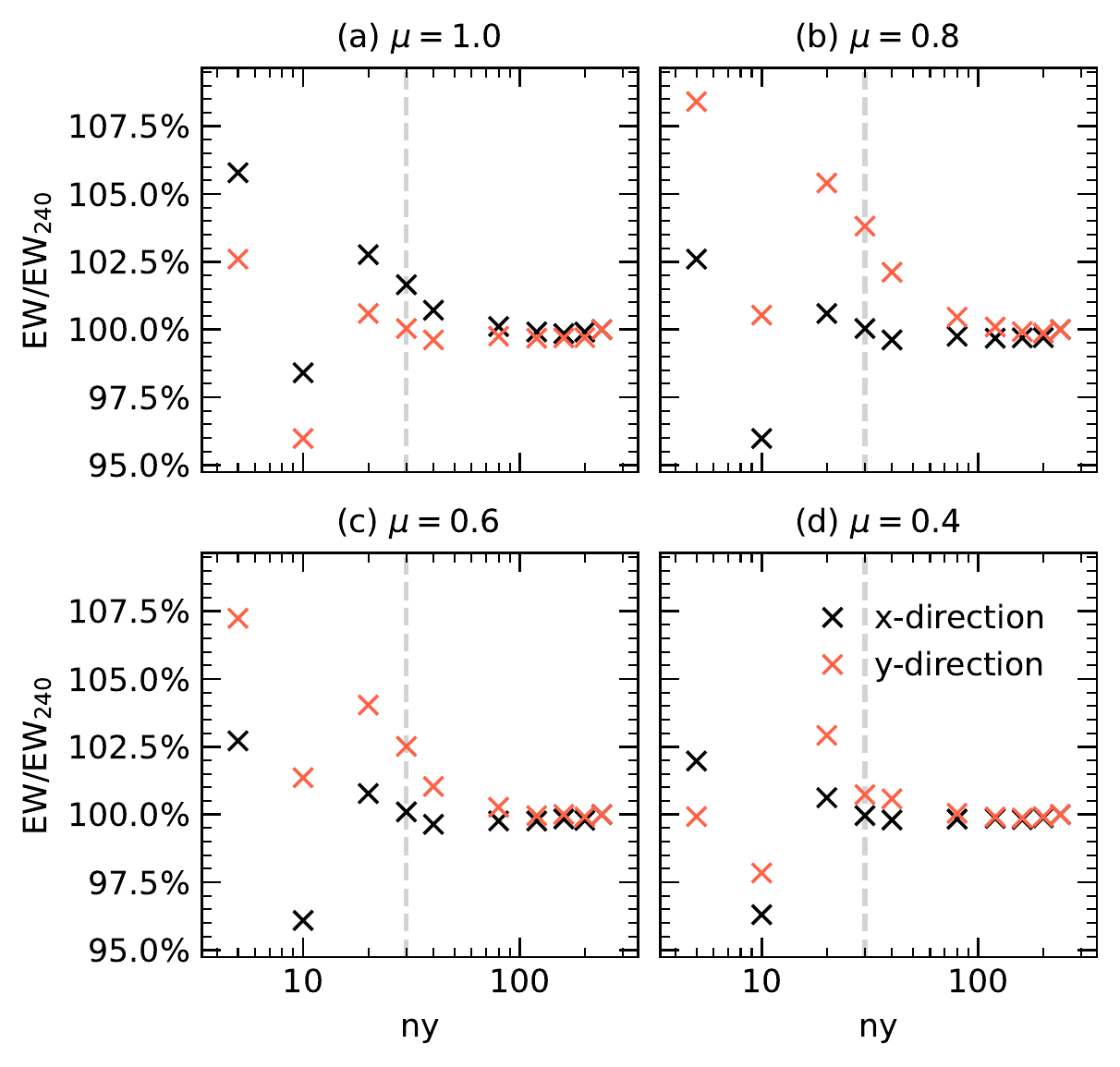}
\caption{Influence of the geometrical resolution (nx,ny) of 3D \texttt{STAGGER} model on the measured equivalent widths. Here nx is kept constant at 5. ny varies from 5 to 240, where 240x240 corresponds to the full resolution of the model. The vertical dashed line shows our default geometric resolution (30,30) adopted in 3D NLTE radiative transfer calculations.}
\label{fig:restest}
\end{figure}
We find that the horizontal resolution of (x,y)$ = (30,30)$ offers a reasonable compromise between the computational expense of 3D NLTE modelling and the physical realism, leading to a small ($\sim 1 - 2 \%$) systematic error in the line strength compared to that obtained the full geometric setup. To correct for this small bias, we extract the geometric resolution correction (Table \ref{tab:errors}) by comparing the EWs of lines computed at our nominal resolution and at the highest-possible resolution. This correction is then separately applied to the abundance determined at each $\mu$ angle. The results are shown in Fig. \ref{fig:restest}. The disc-center and limb results are not very sensitive to this effect, but at $\mu=0.8$ and $\mu=0.6$, the EW error caused by using the horizontal resolution of (x,y)$ = (30,30)$ can be as large as $3 \%$. For the disc-center, the abundance correction for the diagnostic O lines amounts to $0.01$ dex.
\begin{table}
\setlength{\tabcolsep}{2pt}
\renewcommand{\footnoterule}{} 
\caption{Errors in O abundance calculations. The errors caused by the chromosphere and limited geometric resolution in 3D RT calculations are strictly systematic and they are directly applied to the resulting  abundances. We note that these errors refer to the estimates at the disc center. See text.}
\label{tab:errors}
\begin{center}
\begin{tabular}{c ccc  cc}
\toprule 
Line & f-value & Ni blend & Observations & Chromosphere  &  Resolution \\
  \Angstrem  & dex & dex & dex & dex & dex \\     
\midrule
7771.940  & $ \pm$ 0.026 & - & $\pm$ 0.015 & +0.007  & +0.010 \\
7774.170  & $\pm$ 0.026 & - & $\pm$ 0.015 & +0.007  & +0.010 \\
7775.390  & $\pm$ 0.026 & - & $\pm$ 0.015 & +0.005  & +0.010 \\
6300.304  & $\pm$ 0.08   & $\pm$ 0.03 & $\pm$ 0.035 & -0.003  & +0.011 \\
\bottomrule        
\end{tabular} 
\end{center}  
\end{table}  
\subsubsection{Influence of the chromosphere}\label{sec:finalchrom}

We do not use the  \texttt{Bifrost} models to derive the absolute solar O abundance (Sect. \ref{sec:atmos}), but only to quantify the influence of the chromosphere, that is, the re-adjustment of the photospheric structure under the influence of the chromosphere,  on the line formation and on the O abundances. This is done by comparing the results calculated with two types of \texttt{Bifrost} models: (a) the full simulation covering the photosphere and the chromosphere, and (b) the strictly photospheric simulation (Sect. \ref{sec:atmos}). Each simulation represents an extended time series, from which we extract $10$ snapshots for the detailed radiative transfer. The average difference between the $\teff$ values of the snapshots is $~10$ K.

Comparing the O abundances derived using the two \texttt{Bifrost} model sequences in 3D NLTE, we find we find that the abundance differences are not large (Table 5). The chromosphere weakens the 777 triplet lines that corresponds to the increase of O abundance by $\sim +0.01$ dex at the disc center and $+0.02$ dex at the limb relative to a pure photospheric simulation. The 6300 [O I] line is relatively unaffected by the chromosphere, and the O abundance inferred from this line is only $0.003$ dex lower. We note that even ignoring the snapshot with the largest $\teff$ difference ($16$ K) does not alter the estimate of chromospheric correction by more than $0.001$ dex.

Whereas the chromospheric correction for the O lines is small, the presence of a chromosphere may have a larger influence on the spectral lines that form higher up in the layers close and above the temperature minimum, such as the Ni 6767 \AA~helioseismology line.
\begin{table}
\renewcommand{\footnoterule}{} 
\setlength{\tabcolsep}{2pt}
\caption{Final 3D NLTE abundance estimates obtained from the 777 nm and 630 nm oxygen lines.}
\label{tab:finalabund1}
\begin{center}
\begin{tabular}{ccc c}
\toprule
       &    777 nm &  630 nm  &  \\
       &           &          &  \\
\midrule
IAG    &   8.739   &   8.771   &  \\
SST    &   8.716   &   8.737   &  \\
Hinode &   -      &   8.748   &  \\    
DST    &   -      &   8.799   &  \\
\bottomrule
\end{tabular}
\end{center}  
\end{table}

\subsection{Final O abundance}\label{sec:finabund}
Table \ref{tab:finalabund1} provides our estimates of the photospheric O abundance computed in 3D NLTE from different observational datasets, where the results were corrected for the chromospheric and spatial resolution effects (Sects. \ref{sec:finalgeom}, \ref{sec:finalchrom}). The individual results obtained in 1D LTE, 1D NLTE, 3D NLTE with different atomic models are provided in Tab. \ref{tab:finalabundind} (Appendix). The EWs of the corresponding lines are tabulated in Tab. \ref{tab:ewLCAOKaulakys} (Appendix).

To compute the final estimates of O abundance and its uncertainties from different observational data and different atomic models, we proceed as follows. For both 777 and 630 nm lines, we adopt the central values as derived from the IAG data using the LCAO$+$K model atom of O, because of the superior resolving power of these observations and more consistent estimates of abundances derived from the spatially-resolved spectra taken at different pointings across the solar disc \footnote{We note that no optimisation by hand was involved in the analysis.}. The LCAO$+$K (with Kaulakys data) atomic model, furthermore, provides a flatter CLV abundance slope, compared to the LCAO or QPC collisional data, but also from the perspective of atomic physics, the Kaulakys data appear to be necessary, as they are expected to compensate for the lack of short-range interactions in the quantum-mechanical calculations. We also adopt the $S_{\rm bf}=100$ model atom of Ni, as this NLTE model provides the average of the plausible 3D NLTE Ni blend contributions to the [O~I] feature that corresponds to the average between the extremely high, moderate, and negligible collision rates (Sect. \ref{sec:linforNi}). 

\begin{figure}
\includegraphics[scale=0.7]{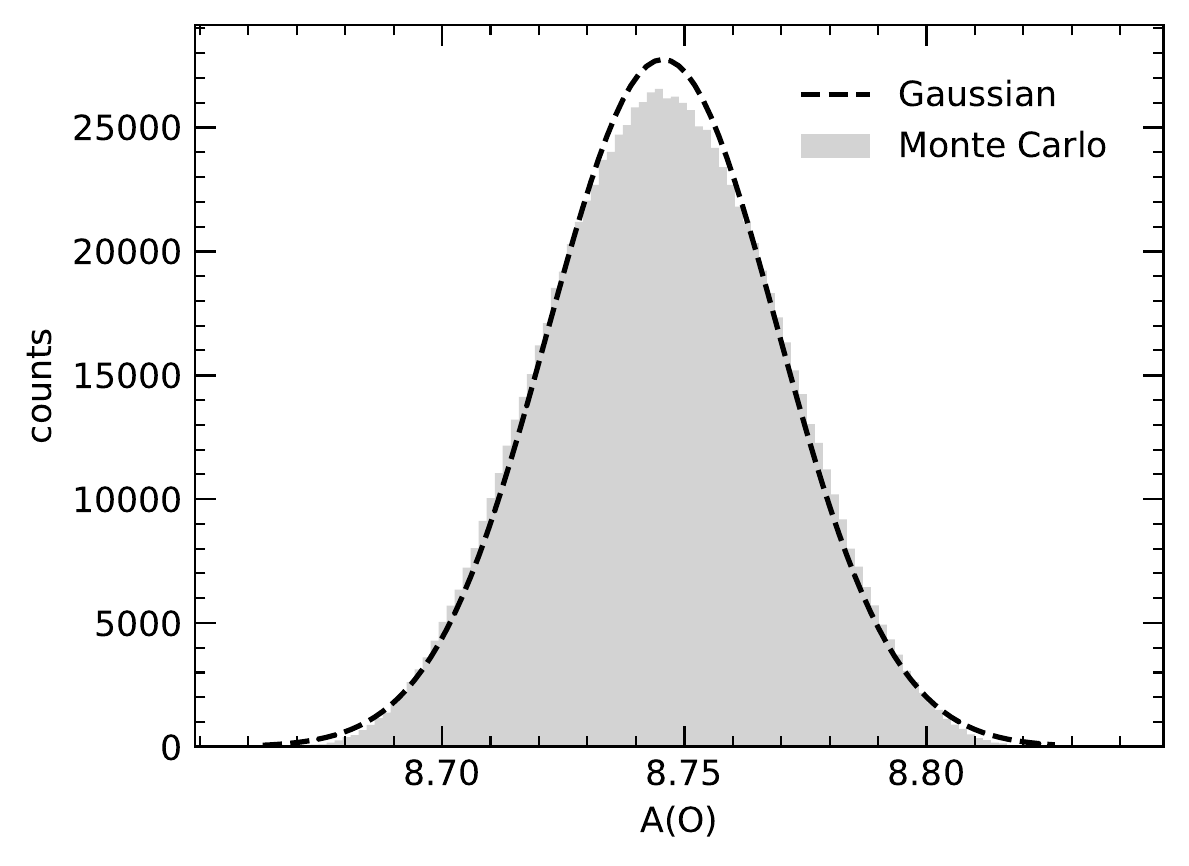}
\caption{Final combined distribution function for the solar photospheric abundance of oxygen. See Sect. \ref{sec:finabund}. A normal distribution is overplotted with a thin line, for comparison.}
\label{fig:pdf}
\end{figure}

To estimate the final combined uncertainties, we resort to a Monte Carlo simulation and make use of the central limit theorem, which states that the normalised sum of random uncorrelated variables closely follows a normal distribution, regardless of the shape of the distribution function of individual variables. This is a commonly used approach in case when the shapes of uncertainties are not known, as in our case. The errors associated with the analysis (see Tab. 5) are used to generate three uniform distributions (large samples of 1 million points each), representing (a) the uncertainty of the oscillator strengths ($f$-value), (b) the uncertainty caused by the Ni blend (630 nm line) and that caused by the O$+$H collisional data (777 nm lines), and (c) the uncertainty caused by using different observed spectra. The uncertainty caused by the collisional data is taken to be the difference between the QPC and LCAO$+$K results. The latter is assumed to be one half of the difference between the SST (min) and the IAG (max) results for the 777 nm line, and that between the SST (min) and DST (max) results for the 630 nm line. These errors are co-added and one standard deviation ($1 \sigma$) of the resulting distribution (Fig. \ref{fig:pdf}) is adopted as the final uncertainty. We note that the final combined distribution of errors for the 777 nm line is very close to normal, whereas for the 630 nm line there is a small deviation owing to a large dominant uncertainty source. However, this has no influence on the conclusions.

As a result, our 3D NLTE value derived from the 777 and 630 nm lines is $\logo = 8.739 \pm 0.027$ dex and $\logo = 8.771 \pm 0.053$ dex, respectively. Assuming that the results are independent, we can combine them to obtain the final photospheric O abundance value of $\logo = 8.75 \pm 0.03$ dex. 

We note that if we were to use the old \citet{Hibbert1991} $f$-values and the SST data, our results based on the LCAO$+$K model atom of O would be in excellent agreement with the 3D NLTE estimate by \citet{Amarsi2018} who relied on the same input. Likewise, our 1D NLTE result based on the MARCS model is in agreement with that by \citet{Sitnova2018}. 

\section{Discussion}\label{sec:disc}
Our estimates of O abundances based on the permitted 777 nm triplet lines and on the 630 forbidden [O~I] line are in a good agreement within the respective uncertainties. The triplet lines are, however, sensitive to the input physics of the O model atom, whereas the forbidden line is sensitive to the input physics of the Ni model atom, primarily to the collisional data. Also, the $f$-values carry a significant source of uncertainty, as the direct experimental verification of the transition probabilities for the 777 nm and 630 nm is not possible. These errors are associated with the physical limitations of theory of atomic physics, molecular physics, and collisional dynamics, and, therefore, their true distributions are unknown. Nonetheless, considering all sources of error, it is very plausible that the photospheric O abundance is close to $\logo = 8.75$ dex.

It is interesting to discuss our results in comparison with the other recent estimates of the solar photospheric O abundance. \citet{Amarsi2018} and \citet{Asplund2021} found $\logo = 8.69 \pm 0.03$ dex and $8.69 \pm 0.04$ dex, respectively. The former estimate is based on the 777 nm line, while the latter also includes the 630 nm and selected molecular OH lines. In principle, our results and those by this group are not inconsistent within their combined uncertainties. Our value is in a much better agreement with the values proposed by \citet{Caffau2008}, \citet{Caffau2015}, and \citet{Steffen2015}, who analysed different O lines, including several forbidden [O~I] lines, the 777 triplet, and the near-IR O I features. Another recent study of the solar photospheric O abundance is that by \citet{Cubas2020}. Their analysis is based on a semi-empirical method of determining the abundance by inverting spatially-resolved observed solar spectra. Their estimate, based on the forbidden [O~I] line at 630 nm, is $\logo = 8.80 \pm 0.03$ dex, also consistent with our results for this feature.
\section{Conclusions}\label{sec:concl}
We present a re-analysis of the solar photospheric O abundance using new atomic data for O and different 3D radiation hydro-dynamical models of solar atmospheres. The NLTE model atom includes novel rates of collisions between O and H atoms, computed using the quantum hopping probability current method \citep{Belyaev2019}, increasing the number of \oi~terms with accurate collisional data to $16$. The R-matrix method for atomic scattering calculations \citep[][]{Berrington1978} is employed to compute new photoionisation cross-sections and the rates describing transitions caused by inelastic processes in collisions of O atoms with free electrons. We also present a new comprehensive model atom of Ni, which, for the lack of detailed quantum-mechanical data, is still based on classical formulae for photoionisation and collisional reaction rates. We use the 1D MARCS solar model atmosphere, the 3D \texttt{Stagger} model \citep{Collet2011, Magic2013}, and two version of 3D MHD models computed self-consistently, with and without the chromosphere, using the \texttt{Bifrost} code \citep{Carlsson2016}.

The model atoms are used to performed 1D LTE, 1D NLTE, and 3D NLTE radiative transfer calculations and abundance analysis of O lines in the solar spectrum. We focus on the least blended lines of \oi~at 7771, 7774, 7775 \Angstrem, and we also consider the forbidden \oi~line at 6300 \Angstrem. We also perform 3D NLTE calculations of the \nii~blend at 6300.341 \Angstrem, which contributes about $25 \%$ of the entire absorption feature. The observational data are taken from different sources. Our primary source of data are the new spatially-resolved spectra taken with the IAG FTS instrument. The data have a resolving power of $R \sim 700\,000$, much higher compared to all previously available solar data. Similar to previous studies, we also include the spectra obtained with the ground-based Swedish Solar Telescope \citep{Pereira2009a}, the Domeless Solar Telescope \citep{takeda2019}, and with the space-based Solar Optical Telescope on the HINODE satellite \citep{Caffau2015}.

We obtain the solar O abundance of $\logo = 8.739 \pm~0.027$ dex and $\logo = 8.771 \pm~ 0.053$ dex based on the 777 nm triplet and 630 nm forbidden [O~I] lines, respectively. Our final combined value  is $\logo = 8.75 \pm~0.03$ dex, which is close to the estimate proposed in the compilation of the solar abundances by \citet[][$8.76 \pm 0.07$ dex]{Caffau2008}, but slightly higher than the recent value by \citet[][$8.69 \pm 0.04$ dex]{Asplund2021}. The differences between our and previous results can be explained by the updated transition probabilities for the 777 lines, new higher-resolution observations collected with the IAG FTS facility, 3D NLTE modelling of the Ni blend, and a correction for a systematic bias caused by the chromospheric back-heating. These effects add up to increase the solar abundance by about $+0.05$ dex relative to the value proposed by Asplund et al. In addition, we confirm that using the HINODE space-based data, which were employed by \citet{Caffau2015}, we recover a slightly higher abundance compared to the value obtained by using the SST data chosen by \citet{Asplund2021}. 

Our results thus suggest that the question of whether the solar photospheric O abundance is low or high, that is closer to the values from \citealt{Grevesse1998}, is still open. There are remaining controversies associated with modelling of the permitted 777 nm lines and of the forbidden [O~I] line. First, the large systematic differences between two sources of H-impact collisional data (\citealt{Kaulakys1991} and \citealt{Belyaev2019}) must be resolved. Second, the differences between the new \textit{lower} $f$-values from \citet{Civis2018} and the older values from \citet{Hibbert1991} must be settled. Until then, we recommend to use the average of the values from both sources. Our results suggest that the indirect influence of the chromosphere on the O lines in the solar spectrum is not large. However, it is  desirable to perform new self-consistent 3D R(M)HD simulations of the solar atmosphere, including chromosphere, with state-of-the-art micro-physics and Non-LTE radiative transfer. 
\section*{Acknowledgements}
We are grateful to the anonymous referee for the helpful remarks and suggestions on the manuscript. We thank Elisabetta Caffau for providing the spectra taken with Solar Optical Telescope on the Hinode satellite. We are grateful to Martin Asplund and Hans-Guenther Ludwig for valuable discussions and comments on the manuscript. SAY, YVV and AKB gratefully acknowledge support from the Ministry of Education (the Russian Federation), project FSZN-2020-0026. MB and ES are supported through the Lise Meitner grant from the Max Planck Society. We acknowledge support by the Collaborative Research centre SFB 881 (projects A5, A10), Heidelberg University, of the Deutsche Forschungsgemeinschaft (DFG, German Research Foundation). ME is supported through the SPP 1992 \emph{Exoplanet Diversity} grant RE 1664/17-1. This project has received funding from the European Research Council (ERC) under the European Union’s Horizon 2020 research and innovation programme (Grant agreement No. 949173).

%%%%%%%%%%%%%%%%%%%%%%%%%%%%%%%%%%%%%%%%%%%%%%%%%%
\section*{Data Availability}\label{sec:dataavail}
The O model atoms (QPC, LCAO, LCAO$+$K) and the error map representing the  blends around the 630.0 nm O$+$Ni feature are available at \url{https://keeper.mpdl.mpg.de/d/da74c98c1c9c4477b67f/}
 


\bibliographystyle{mnras}
\bibliography{references} % if your bibtex file is called example.bib

\appendix

\section{Appendix}

\begin{figure}
\includegraphics[scale=0.47]{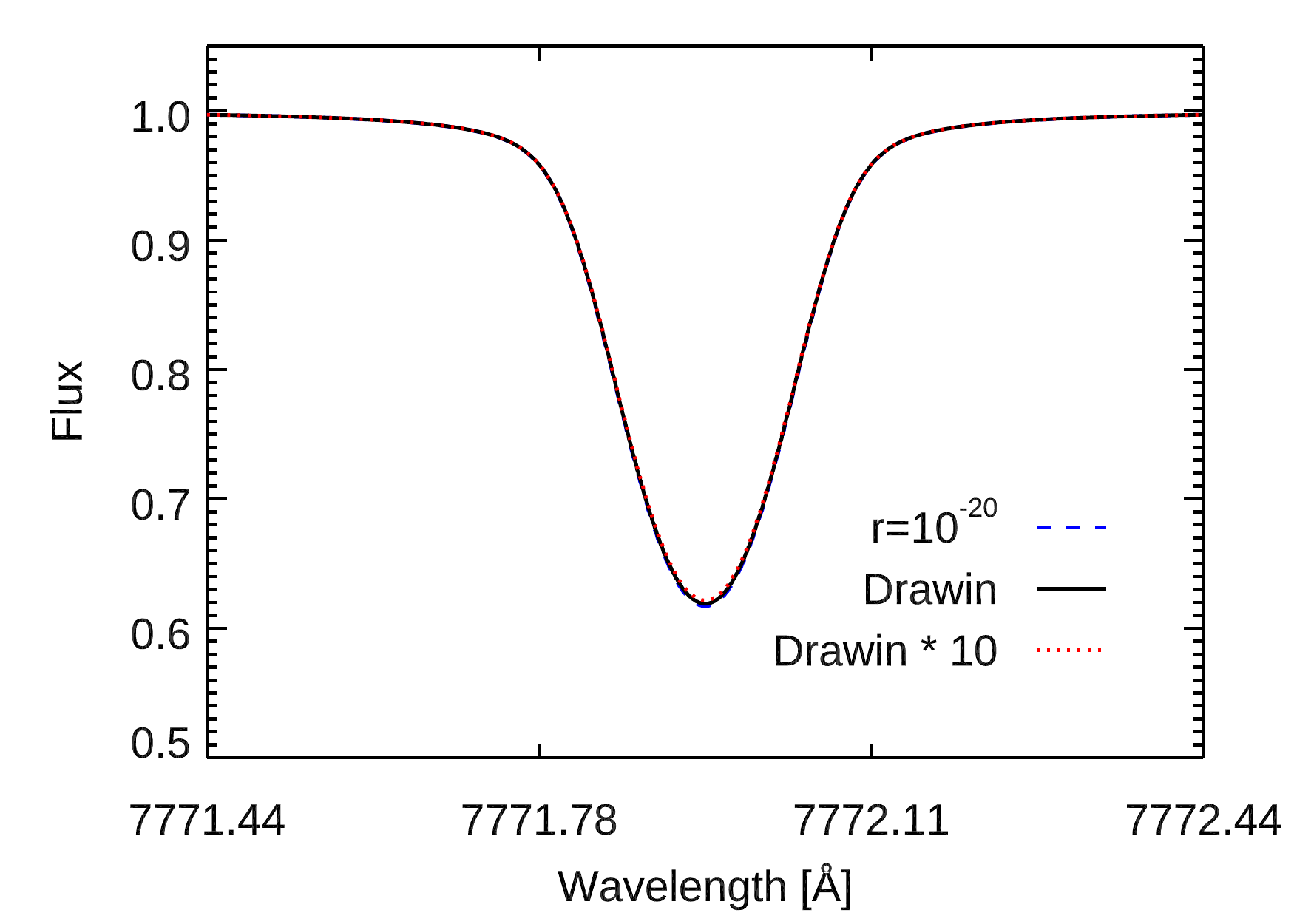}
\caption{Profiles of the 7771 \AA\ line computed in NLTE using three different atomic models with the collisional rate coefficients for the upper 100 states uniformly set to $10^{-20}$, assumed to follow the Drawin's recipe, and scaled by a factor of 10 relative to Drawin.}
\label{fig:DrScal}
\end{figure}

\begin{figure*}
\includegraphics[scale=0.7]{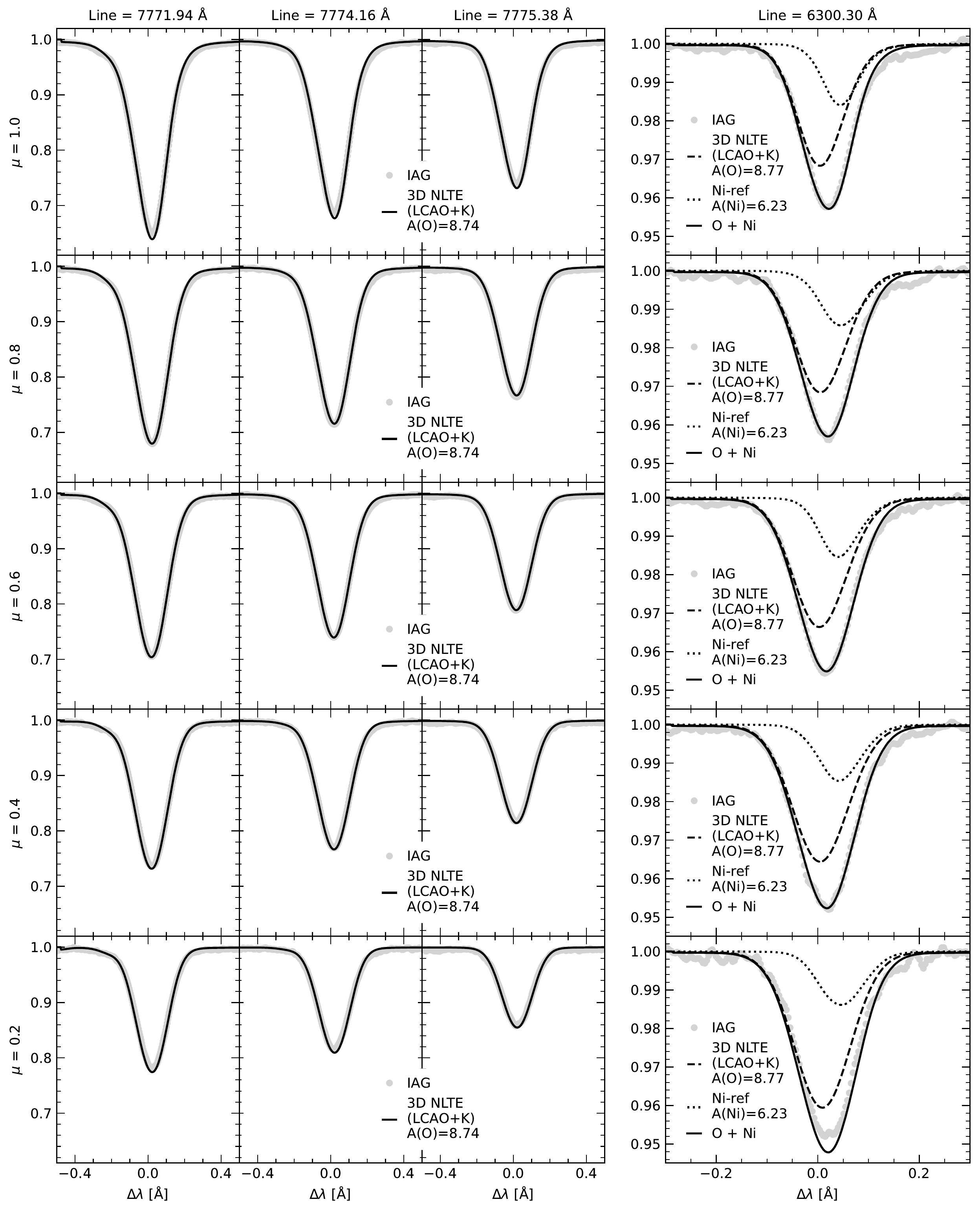}
\caption{3D NLTE profiles of the 777 nm and 630 nm lines computed using the \texttt{Stagger} 3D model atmosphere and the LCAO + Kaulakys model atom, compared with the spatially-resolved IAG observations of the Sun for different $\mu$ angles (top to bottom).}
\label{fig:CLVStaggerLCAO}
\end{figure*}
\begin{figure*}
\includegraphics[scale=0.7]{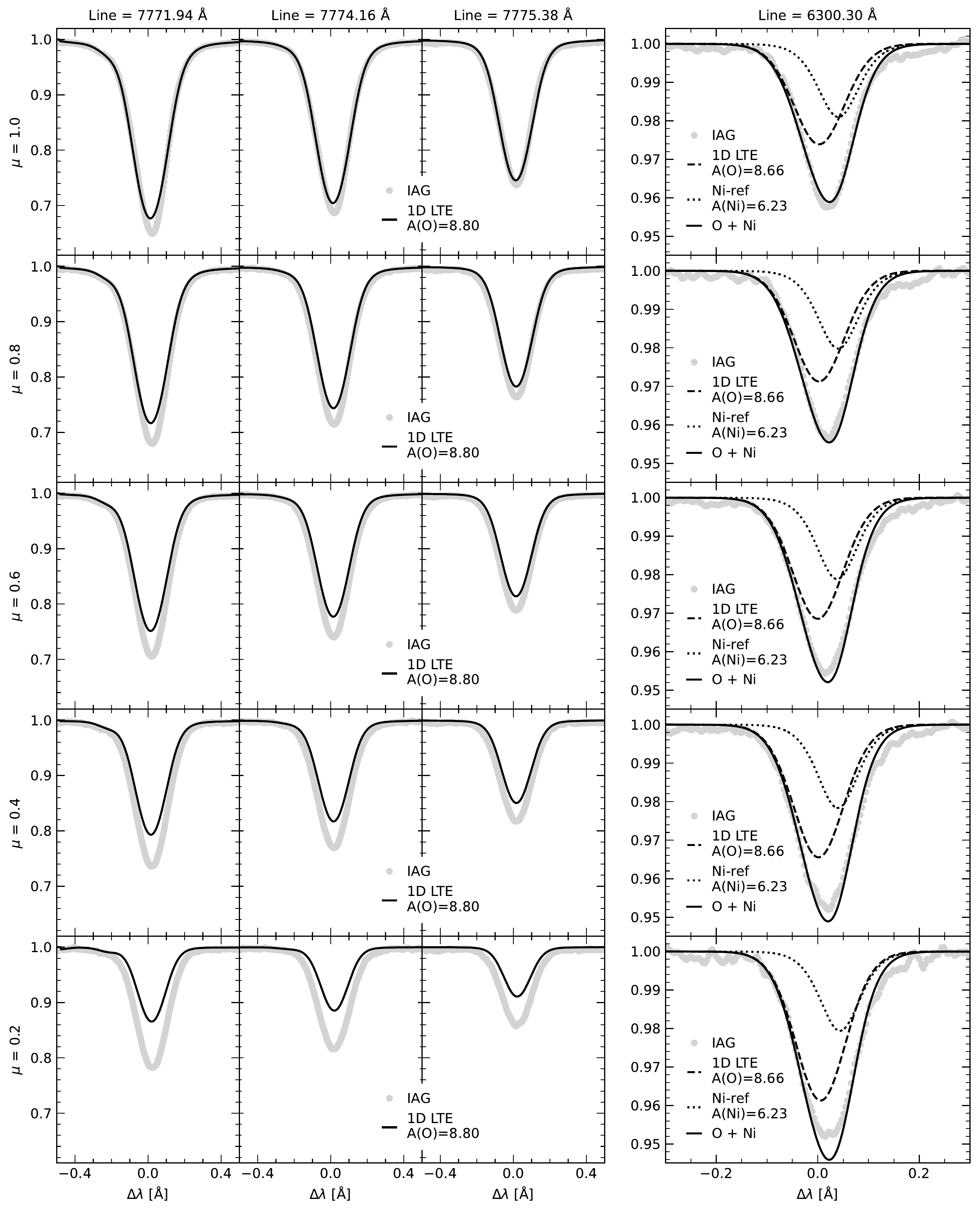}
\caption{1D LTE profiles of the 777 nm and 630 nm lines computed using the MARCS 1D model atmosphere and the LCAO + Kaulakys model atom, compared with the spatially-resolved IAG observations of the Sun for different $\mu$ angles (top to bottom).}
\label{fig:CLV1dlte}
\end{figure*}
\begin{figure*}
\hbox{
\includegraphics[scale=0.7]{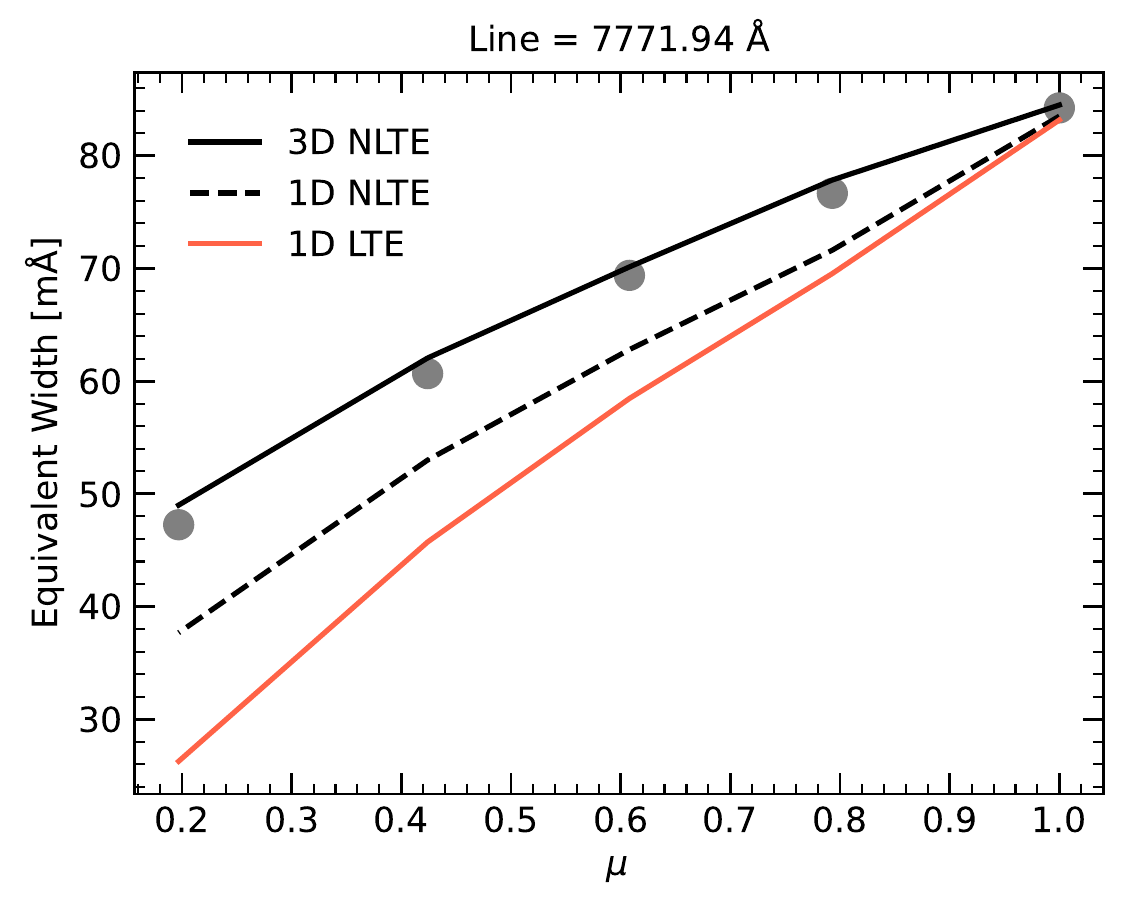}
\includegraphics[scale=0.7]{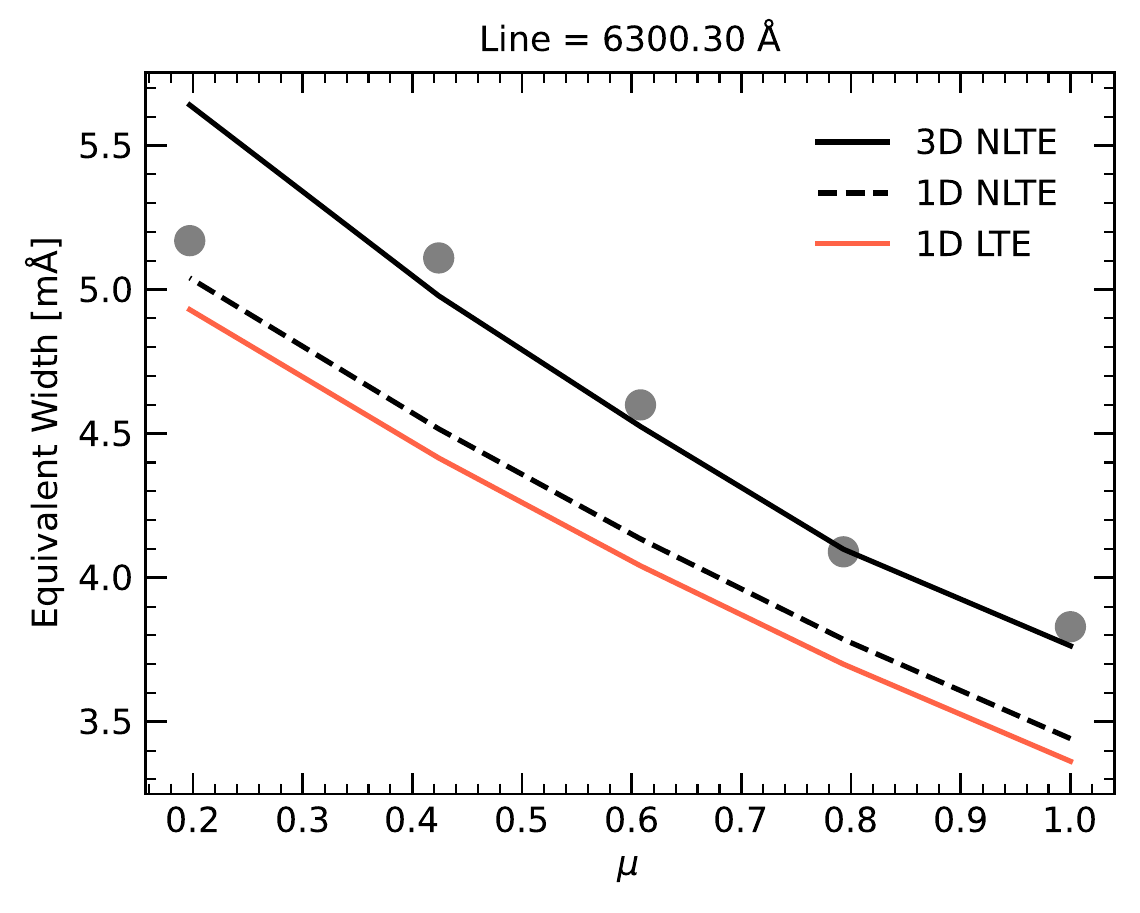}
}
\caption{Comparison of observed EWs of the 7771.94 and 6300.3 \Angstrem~features with the predictions of 1D LTE, 1D NLTE, 3D NLTE models. We note that the O abundances are determined using the more correct full profile fitting method, not from the line EWs.}
\label{fig:CLVall1}
\end{figure*}

\begin{figure}
\includegraphics[scale=0.7]{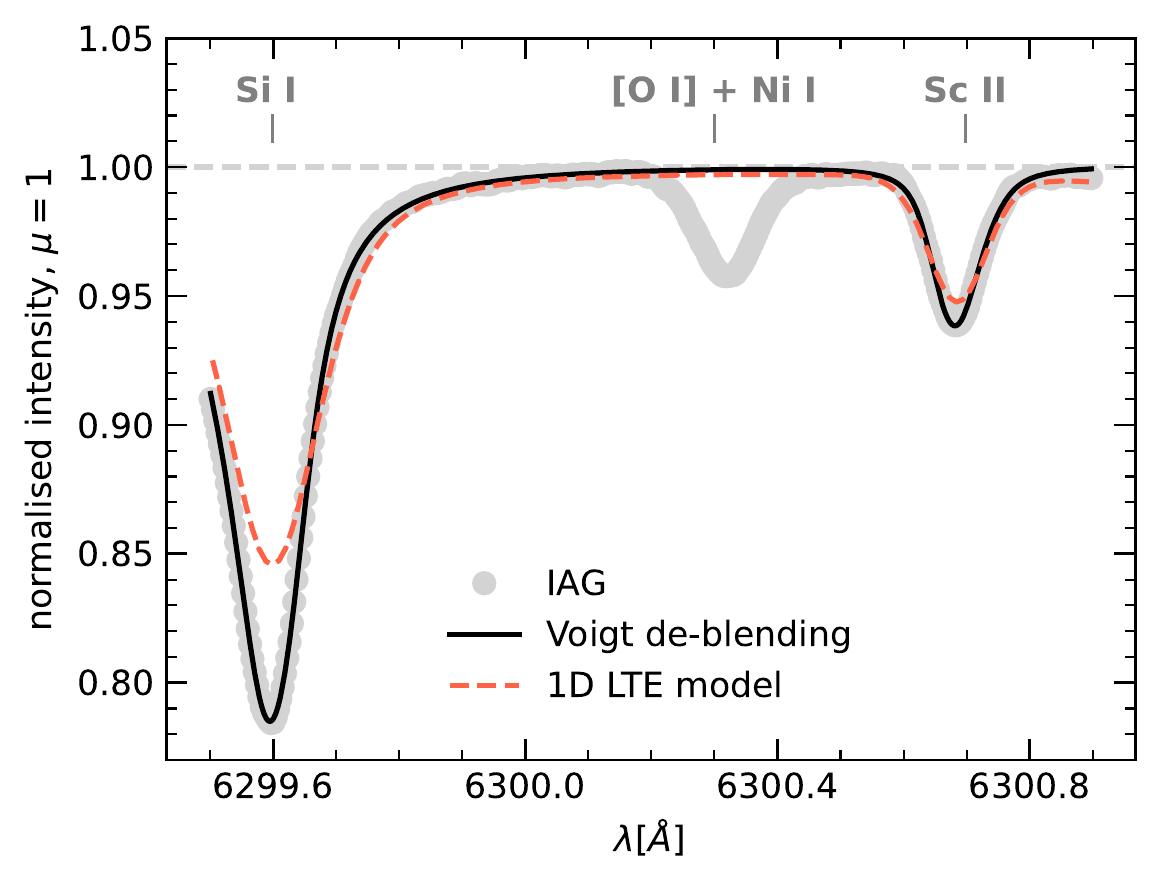}
\caption{Spectral lines surrounding the oxygen + nickel blend at 630 nm at the disc center $\mu=1.0$.}
\label{fig:blends}
\end{figure}

\begin{figure}
    \centering
    \includegraphics[scale=0.7]{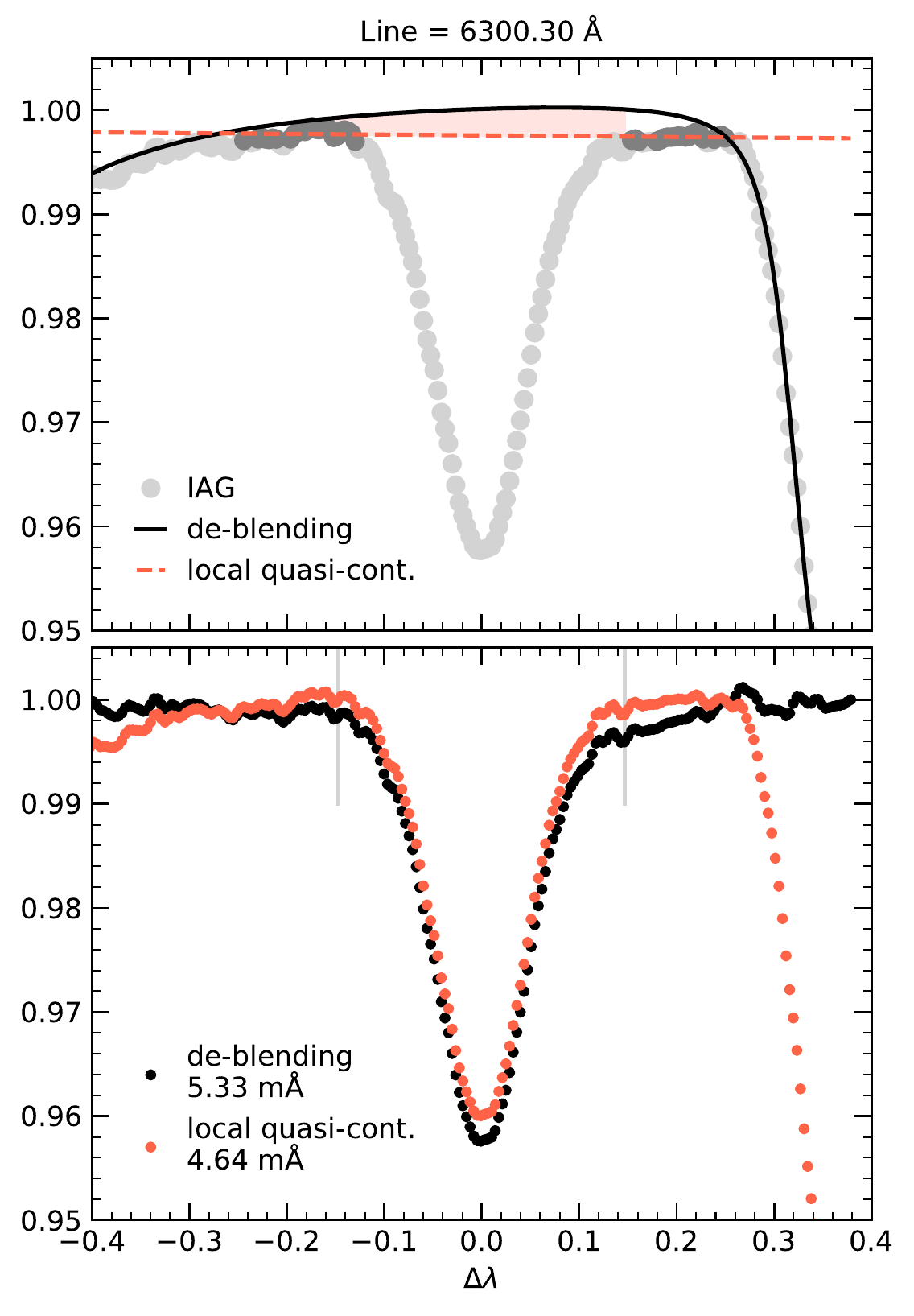}
    \caption{Top panel shows the Voigt profiles used to de-blend of the 630 nm [O~I]$+$Ni feature. In addition, we show the classical technique, in which a local quasi-continuum is fit directly through the local maxima of the relative flux (here, just for demonstration we employ the dark grey points). The grey lines in the bottom panel depict the integration limits of the equivalent widths stated in the legend. These EW limits are used for demonstration only, but not for the  abundance estimate (see Sect. 4.6).}
    \label{fig:deblending}
\end{figure}

\begin{table*}
\renewcommand{\footnoterule}{} 
\caption{Individual abundance estimates for different diagnostic lines of O (777 triplet and 630 forbidden features). We note that the modelling of the 630 nm Ni I line is fully consistent with the modelling of O lines. That is, Ni is computed in 1D LTE if 1D LTE is adopted for O, and in 3D NLTE if 3D NLTE is used for O.}
\label{tab:finalabundind}
\begin{center}
\begin{tabular}{llllllll}
\toprule
       &                 & 1D LTE & 3D LTE & 1D NLTE  & 3D NLTE  & 3D NLTE & 3D NLTE \\
       &                 &        &        & (LCAO+K) & (LCAO)   & (QPC)   & (LCAO+K) \\
Spectrum & Line &        &        &                   &                   &                &                 \\
\midrule
IAG &  777           &  8.803 &  8.876 &             8.691 &           8.643 &          8.671 &             8.739 \\
     & 630 (Ni-max)  &        &        &                   &           8.795 &          8.795 &             8.795 \\
     & 630 (Ni-ref)  &  8.661 &  8.743 &             8.671 &           8.771 &          8.771 &             8.771 \\
     & 630 (Ni-min)  &        &        &                   &           8.743 &          8.743 &             8.743 \\
      & & & & & & & \\ 
SST &  777           &  8.784 &  8.856 &             8.671 &           8.621 &          8.648 &             8.716 \\
     & 630 (Ni-max)  &        &        &                   &           8.763 &          8.763 &             8.763 \\
     & 630 (Ni-ref)  &  8.622 &  8.705 &             8.632 &           8.737 &          8.737 &             8.737 \\
     & 630 (Ni-min)  &        &        &                   &           8.706 &          8.706 &             8.706 \\
      & & & & & & & \\        
Hinode &  777        &        &        &                   &                 &                &                   \\
     & 630 (Ni-max)  &        &        &                   &           8.773 &          8.773 &             8.773 \\
     & 630 (Ni-ref)  &  8.633 &  8.719 &             8.643 &           8.748 &          8.748 &             8.748 \\
     & 630 (Ni-min)  &        &        &                   &           8.720 &          8.720 &             8.720 \\
      & & & & & & & \\        
DST &  777           &        &        &                   &                 &                &                   \\
     & 630 (Ni-max)  &        &        &                   &           8.822 &          8.822 &             8.822 \\
     & 630 (Ni-ref)  &  8.699 &  8.774 &             8.708 &           8.799 &          8.799 &             8.799 \\
     & 630 (Ni-min)  &        &        &                   &           8.774 &          8.774 &             8.774 \\

      & & & & & & & \\    
Neckel &  777        &  8.793 &  8.866 &             8.682 &           8.635 &          8.662 &             8.729 \\
     & 630 (Ni-max)  &        &        &                   &           8.788 &          8.788 &             8.788 \\
     & 630 (Ni-ref)  &  8.652 &  8.734 &             8.662 &           8.763 &          8.763 &             8.763 \\
     & 630 (Ni-min)  &        &        &                   &           8.734 &          8.734 &             8.734 \\
      & & & & & & & \\    
KPNO &  777          &  8.797 &  8.869 &             8.682 &           8.632 &          8.660 &             8.728 \\
     & 630 (Ni-max)  &        &        &                   &           8.803 &          8.803 &             8.803 \\
     & 630 (Ni-ref)  &  8.671 &  8.751 &             8.680 &           8.779 &          8.779 &             8.779 \\
     & 630 (Ni-min)  &        &        &                   &           8.751 &          8.751 &             8.751 \\
\bottomrule
\end{tabular}
\end{center}  
\end{table*}

\begin{table*}
\caption{EWs estimates of spectral lines obtained using 3D NLTE line profiles of the LCAO + Kaulakys model atom. All equivalent widths in mÅ.}
\label{tab:ewLCAOKaulakys}
\begin{center}
\begin{tabular}{lrrrrrrr}
\hline 
Specie & line [Å] &  $\mu$ = 1.0 &  $\mu$ = 0.8 &  $\mu$ = 0.6 &  $\mu$ = 0.4 &  $\mu$ = 0.2 \\
\midrule
O I    & 7771.94 &  84.25 &  76.67 &  69.39 &  60.67 &  47.26 \\
O I    & 7774.17 &  74.48 &  67.55 &  60.70 &  52.52 &  40.28 \\
O I    & 7775.39 &  59.26 &  53.30 &  47.15 &  39.88 &  29.49 \\
O I    & 6300.30 &   3.83 &   4.09 &   4.60 &   5.11 &   5.17 \\
Ni I   & 6300.34 &   1.49 &   1.56 &   1.66 &   1.72 &   1.72 \\
\hline 
\hline 
\end{tabular}  
\end{center}
\end{table*}
%
%

% Don't change these lines
\bsp	% typesetting comment
\label{lastpage}
\end{document}